\documentclass[authoryear, 5p, preprint]{elsarticle}

\usepackage[utf8]{inputenc} %
\usepackage[T1]{fontenc}    %
\usepackage{lmodern}
\usepackage[hidelinks]{hyperref}       %
\usepackage{url}            %
\usepackage{booktabs}       %
\usepackage{amsfonts}       %
\usepackage{nicefrac}       %
\usepackage{microtype}      %
\usepackage{fullpage}
\usepackage{lineno}
\usepackage{natbib}
\usepackage{pifont}
\usepackage{amsmath,amssymb}
\usepackage{graphicx}
\usepackage{subcaption}
\usepackage{multicol,graphicx,xcolor,multirow, array}
\usepackage[normalem]{ulem}
\usepackage{mathtools}
\usepackage{accents} %
\usepackage[export]{adjustbox} %
\usepackage{tikz}
\usetikzlibrary{shapes.misc, positioning}
\usetikzlibrary{matrix, calc, positioning, arrows,shapes,backgrounds, arrows.meta, quotes}
\usepackage{gensymb,stmaryrd,mathtools,textcomp,xspace,grffile}
\usetikzlibrary{shapes.geometric,fit,matrix,positioning,shapes.multipart}
\usetikzlibrary{decorations.markings}
\usetikzlibrary{shadows,decorations.pathreplacing,fadings}
\pgfdeclarelayer{background}
\pgfsetlayers{background,main}
\usetikzlibrary{topaths, calc,3d}
\usetikzlibrary{fit}

\def\bl{\pmb{\ell}}
\def\bL{\mathbf{L}}
\def \myfigure {Fig.}
\def \myfigures {Figs.}
\def \myequation {Eqn.}

\newlength{\dhatheight}
\newcommand{\dbhat}[1]{%
    \settoheight{\dhatheight}{\ensuremath{\hat{#1}}}%
    \addtolength{\dhatheight}{-0.35ex}%
    \hat{\vphantom{\rule{1pt}{\dhatheight}}%
    \smash{\hat{#1}}}}
    
\newcommand{\dbhatss}[1]{%
    \settoheight{\dhatheight}{\ensuremath{\hat{#1}}}%
    \addtolength{\dhatheight}{-0.8ex}%
    \hat{\vphantom{\rule{1pt}{\dhatheight}}%
    \smash{\hat{#1}}}}
    
\newcommand{\dbhatssk}[1]{%
    \settoheight{\dhatheight}{\ensuremath{\hat{#1}}}%
    \addtolength{\dhatheight}{-0.7ex}%
    \hat{\vphantom{\rule{1pt}{\dhatheight}}%
    \smash{\hat{#1}}}}

\cortext[cor1]{Corresponding author}

\def \FigureEkspec {

\begin{figure}
\begin{subfigure}{\linewidth}
\centering
  \includegraphics[width=\linewidth]{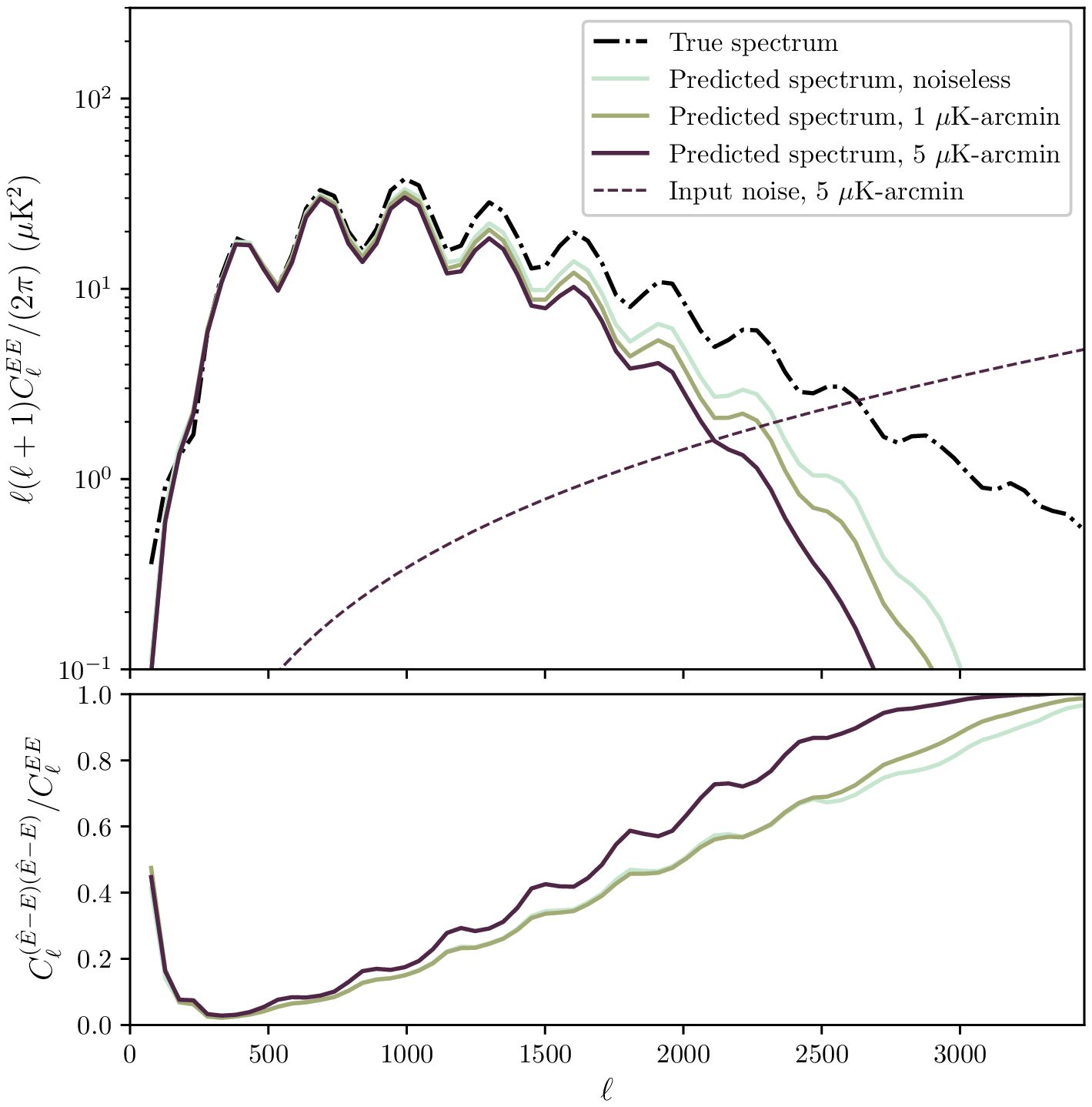}
\caption{$E$ spectra. For comparison, we also show the spectrum of the 5 $\mu$K-arcmin white noise applied to the inputs.} \label{fig:Espec}
\end{subfigure}
\hfill
\begin{subfigure}{\linewidth}
\centering
  \includegraphics[width=\linewidth]{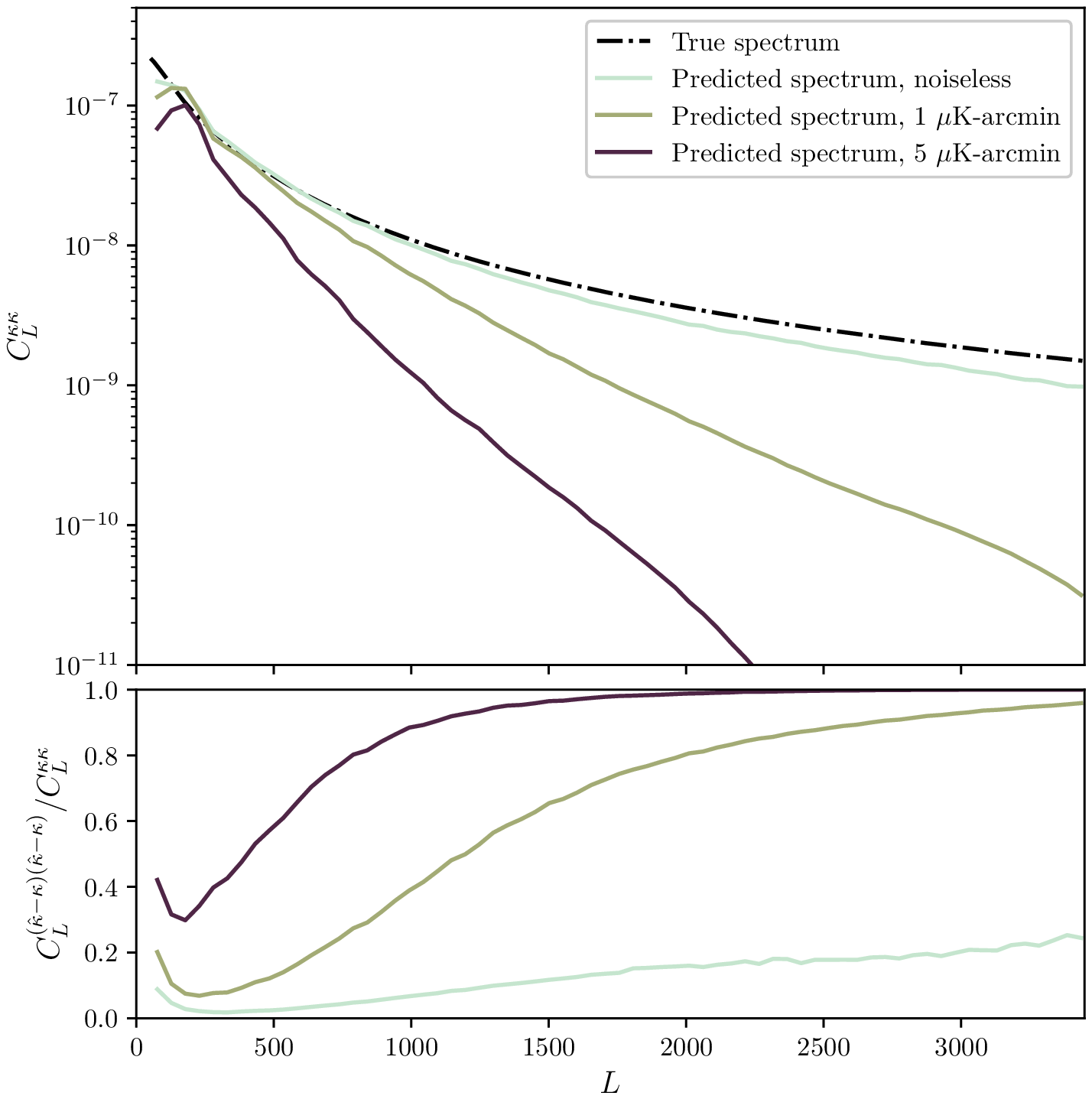}
\caption{$\kappa$ spectra.} \label{fig:kspec}
\end{subfigure}
\caption{In the top panels, we see the power spectra of true and recovered $E$ and $\kappa$ maps, averaged over all realizations in the test set. Note that the recovered spectra degrade as noise is increased in the input maps. In the noiseless case, $\kappa$ recovery is more successful than that of $E$ across a larger $L$ range, as we had anticipated in \myfigure~\ref{fig:kmaps}. However noise has a much larger effect on $\kappa$ recovery, strongly degrading its quality while $E$ recovery stays qualitatively similar.
Bottom panels show the ratio of the difference-map auto-spectrum to the input map auto-spectrum. 
The difference maps constitute the difference between the network-predicted outputs and the input maps for each noise level. 
} \label{fig:spectra}
\end{figure}

}

\def \Figuresmallerimages {

\begin{figure}
\begin{subfigure}[t]{\linewidth}
\centering
  \includegraphics[width=\linewidth]{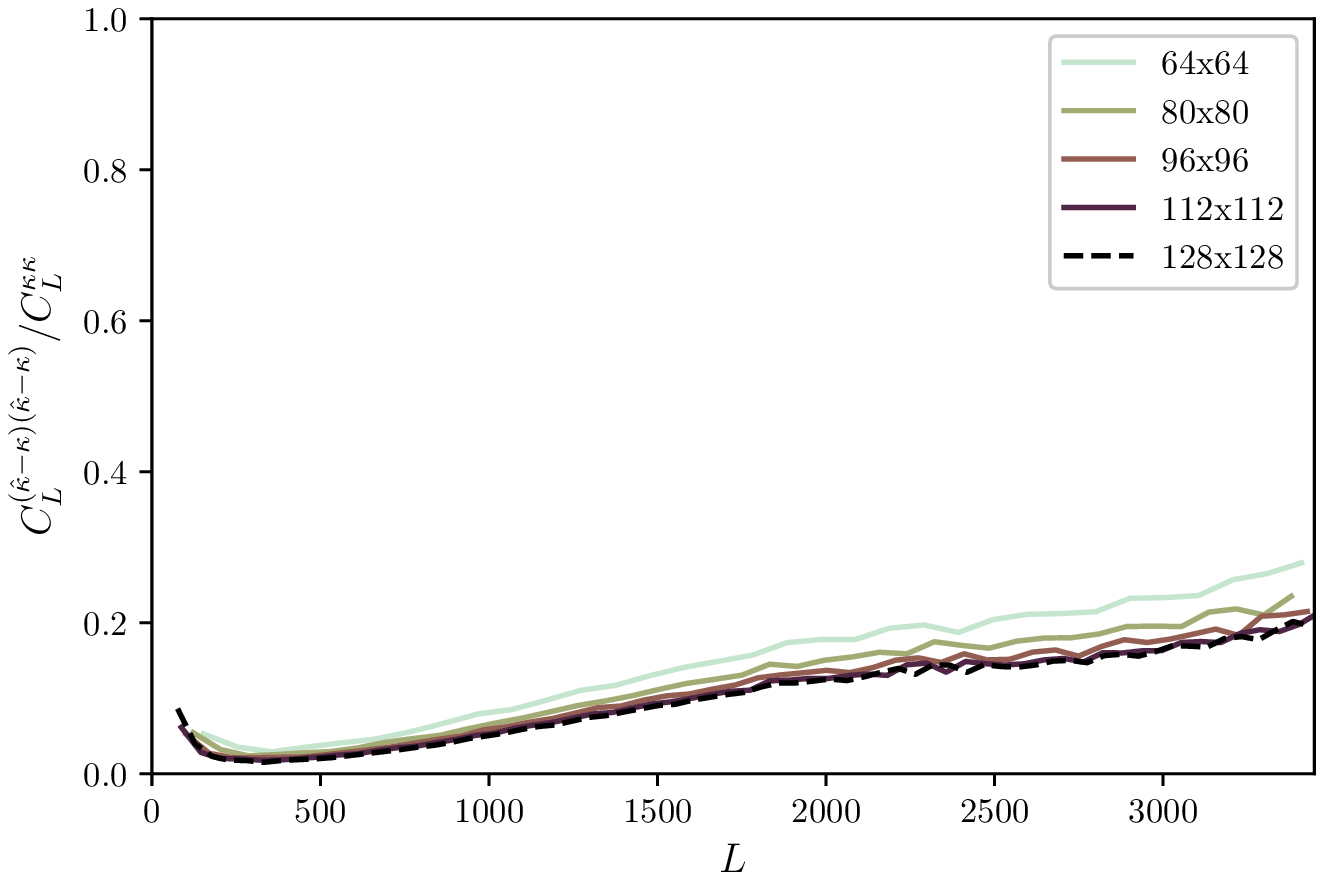}
\caption{Ratio between the spectrum of the difference-maps and spectrum of the true maps for different cutout sizes.} \label{fig:smaller_zoomout}
\end{subfigure}
\hfill
\begin{subfigure}[t]{\linewidth}
\centering
  \includegraphics[width=\linewidth]{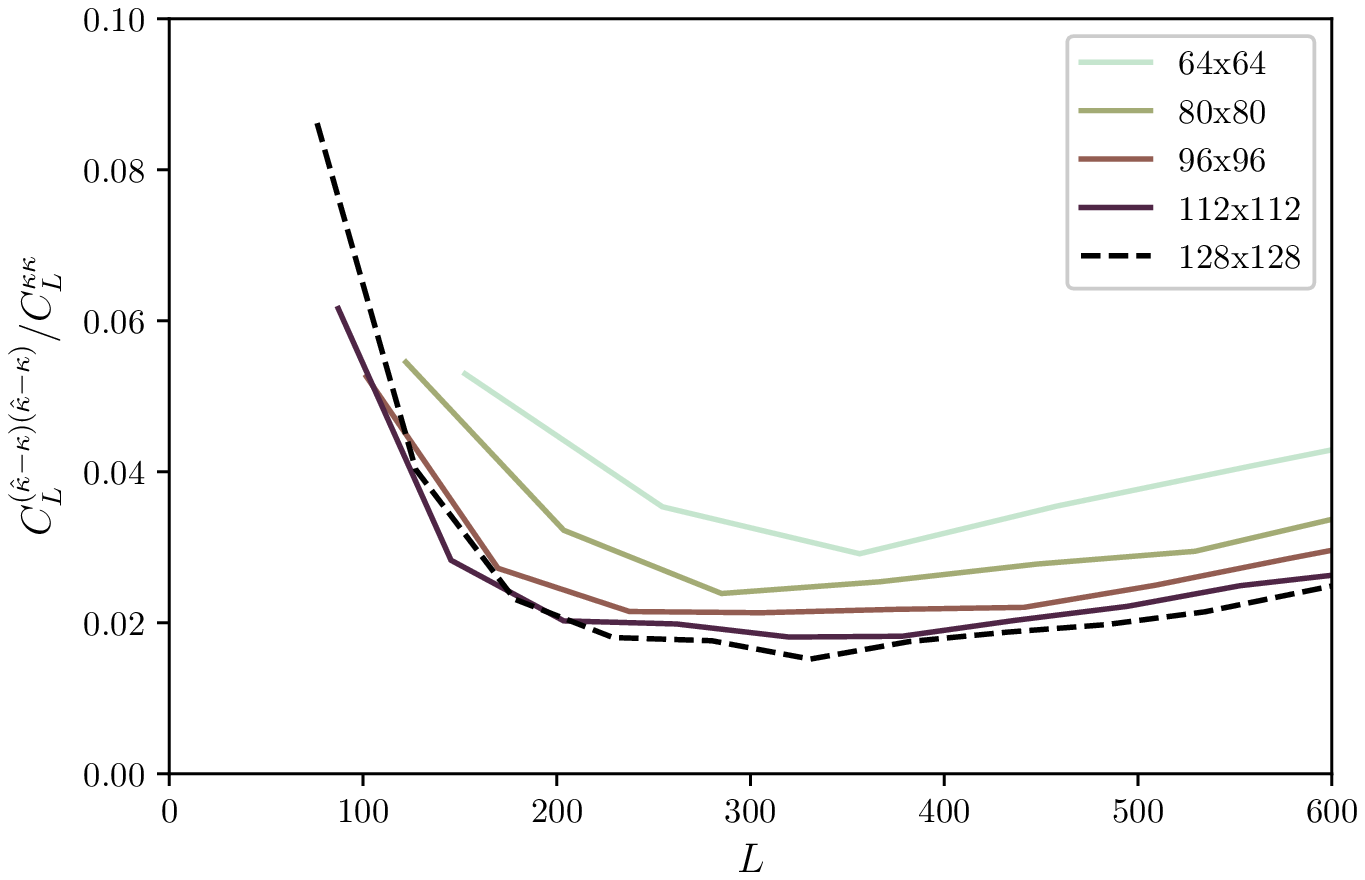}
\caption{Same ratio, zoomed into the low $L$ region and capped at 10\%.} \label{fig:smaller_zoomin}
\end{subfigure}
\caption{We vary the cutout size and look at the ratio plotted in the bottom panel of \myfigure~\ref{fig:kspec}, each line corresponding to a new network trained on images of that size. We can see that performance degrades overall as the cutouts become smaller, as might be expected since there is less information contained in the inputs. On the bottom we zoom in on the low-$L$ region, showing that the uptick in relative difference occurs at larger $L$ for smaller images. This suggests that low-$L$ performance could be improved by working on a larger patch of sky.} \label{fig:smallerimages}
\end{figure}

}

\def \FigureclkkQEIMvsResUNet {

\begin{figure}
\begin{subfigure}[t]{\linewidth}
\centering
  \includegraphics[width=\linewidth]{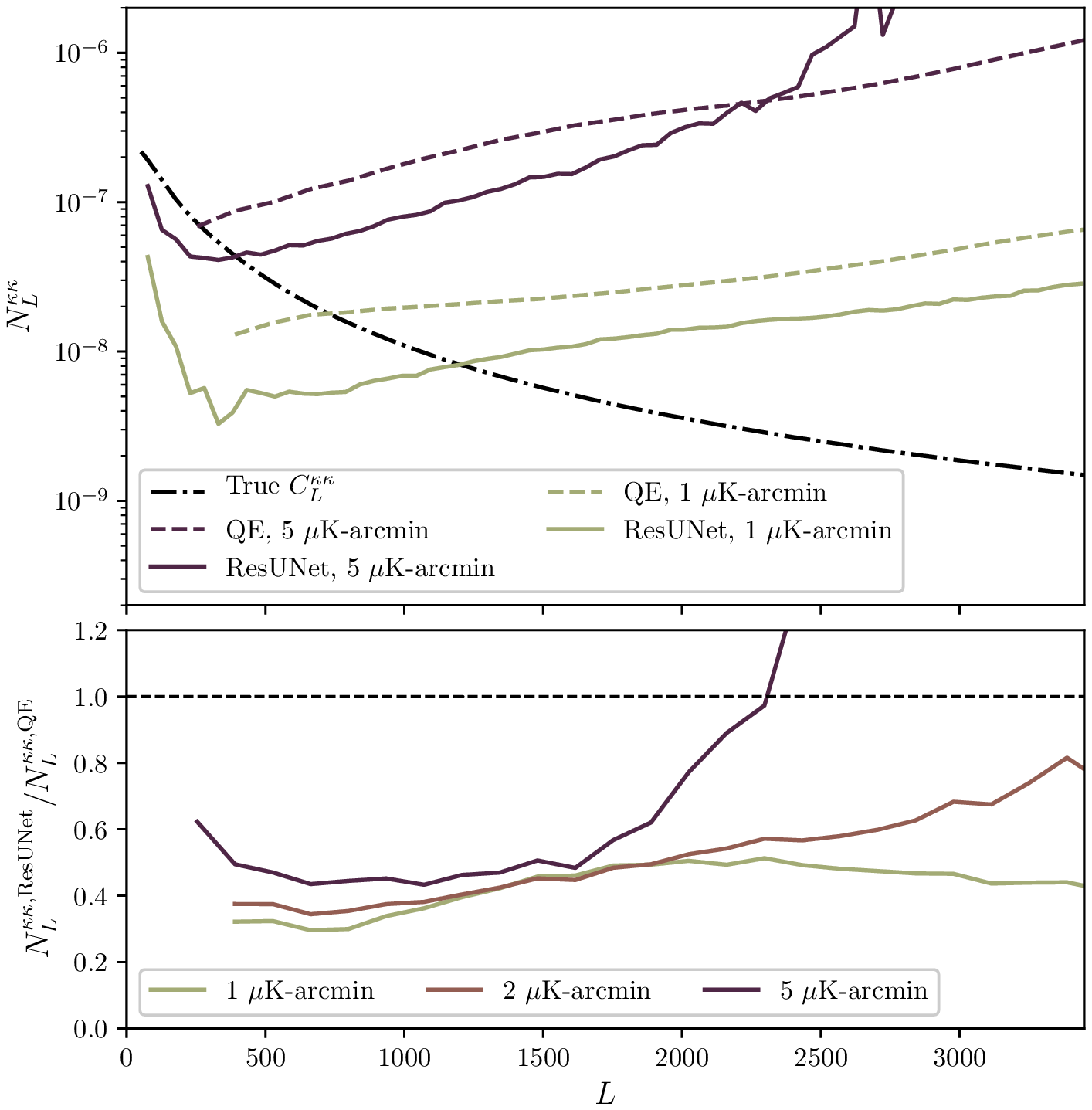}
\caption{To further evaluate the quality of $\kappa$ reconstruction by ResUNets at different input noise levels, we compare the noise spectra from ResUNets to those from quadratic estimators. We see that the results have $50-70\%$ less noise than quadratic estimator reconstructions across a wide range of angular scales $L$. For input noise of 5 $\mu$K-arcmin, performance quickly degrades for $L \gtrsim 2000$.} \label{fig:clkkQEvsResUNet}
\end{subfigure}
\hfill
\begin{subfigure}[t]{\linewidth}
\centering
  \includegraphics[width=\linewidth]{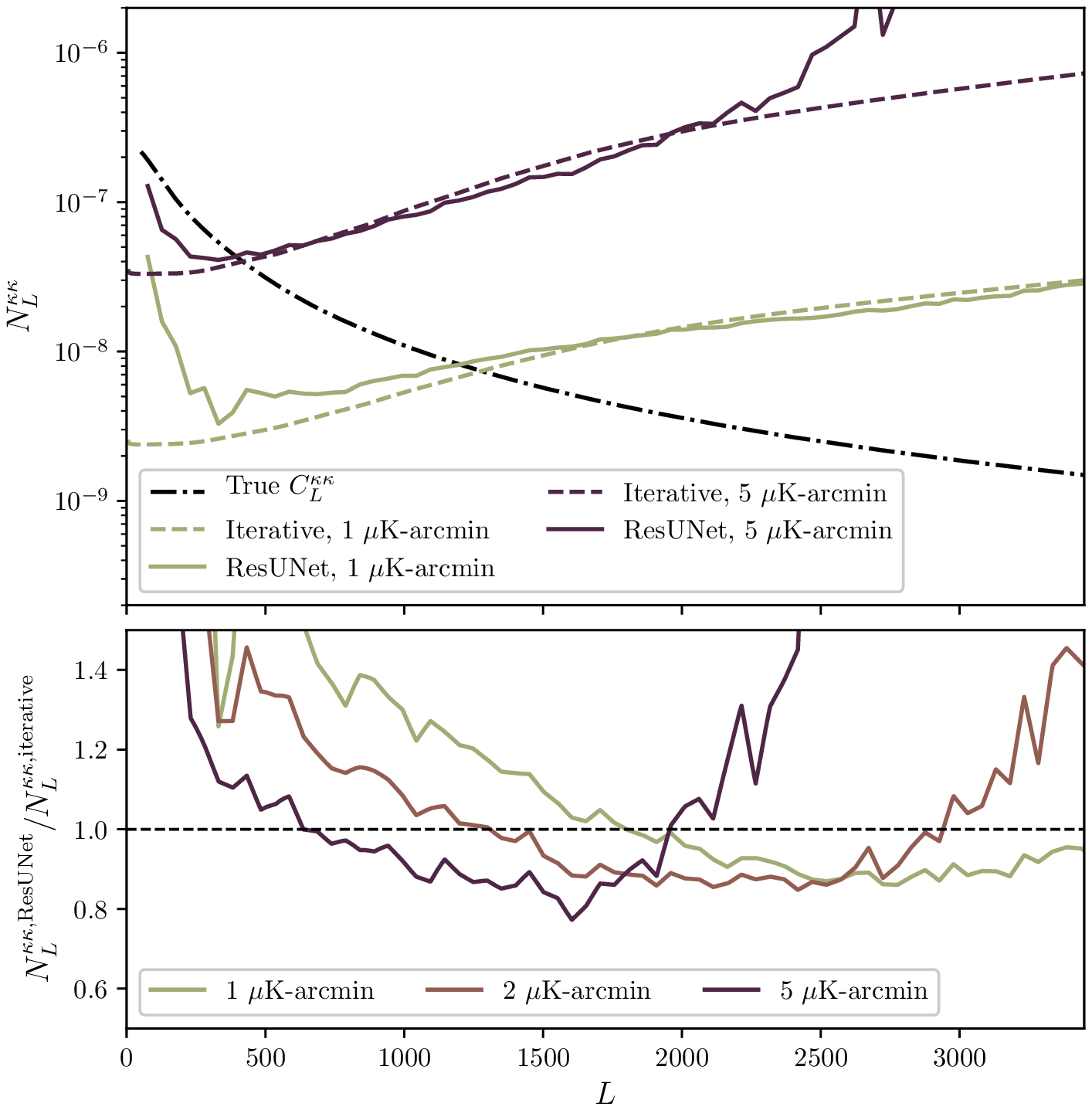}
\caption{We also compare noise spectra of $\kappa$ reconstruction using ResUNets to expected noise levels from iterative estimators (which approach maximum-likelihood results), as in \cite{smith10}. The iterative method noise levels are taken from an $EB$ estimator only. For all noise levels used here, ResUNets and iterative methods have comparable noise levels across a wide range of $L$. Significant performance differences mainly occur for the smallest and largest scales pictured.} \label{fig:clkkIMvsResUNet}
\end{subfigure}
\caption{We compare $\kappa$ reconstruction using ResUNets to current standard methods. The noise spectra are calculated by taking the average spectrum over all realizations in the test set. 
} \label{fig:comparisonstoothermethods}
\end{figure}

}

\def \FigureInputmapsnoisev2 {

\begin{figure*}
\centering
\bgroup
{\setlength{\tabcolsep}{0em}
\def\arraystretch{0.} 
\begin{tabular}{r r r r}
\multicolumn{1}{c}{0 $\mu$K-arcmin} & \multicolumn{1}{c}{1 $\mu$K-arcmin} & \multicolumn{1}{c}{5 $\mu$K-arcmin} & \\
\includegraphics[width=.300242\textwidth]{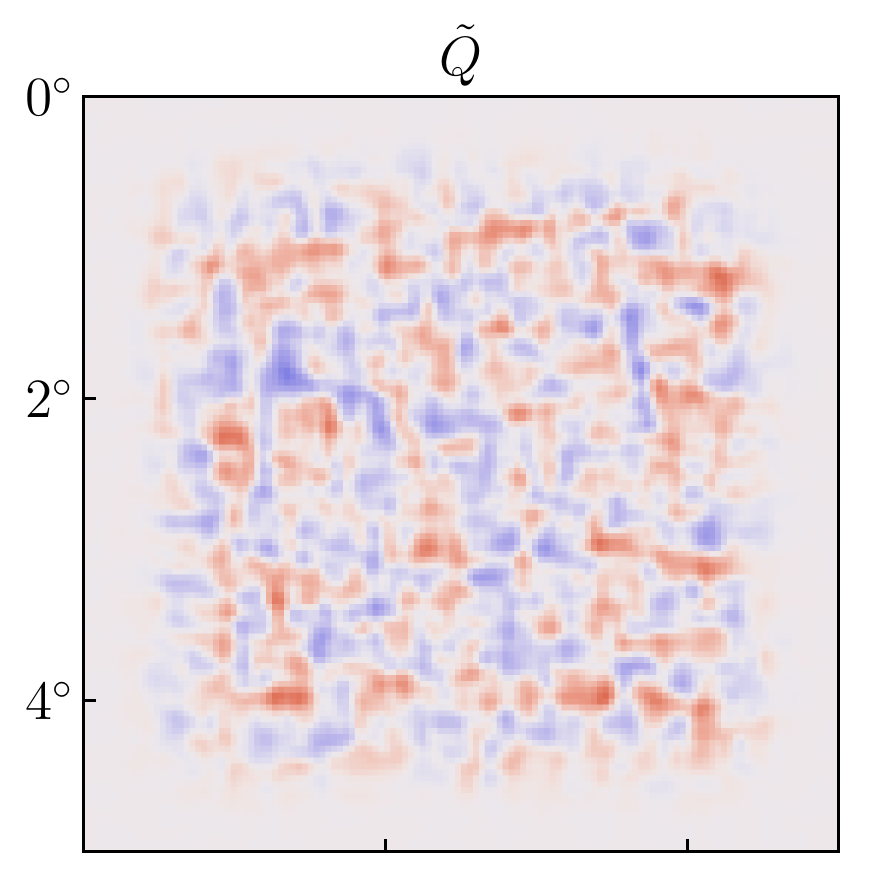}
&
\includegraphics[width=.28\textwidth]{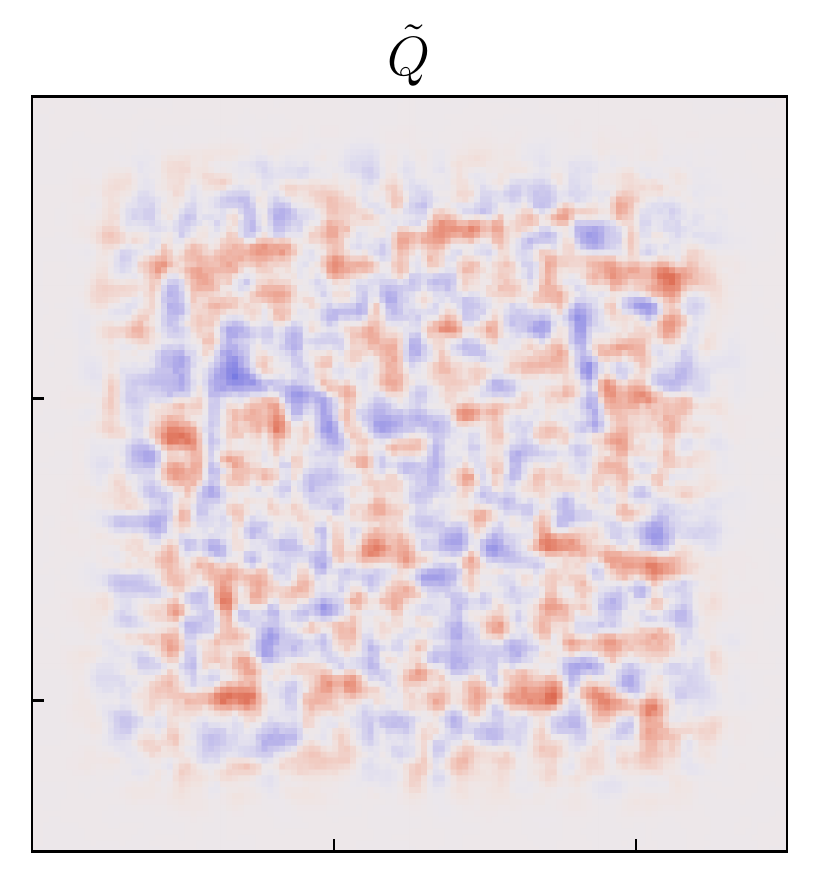}
&
\includegraphics[width=.28\textwidth]{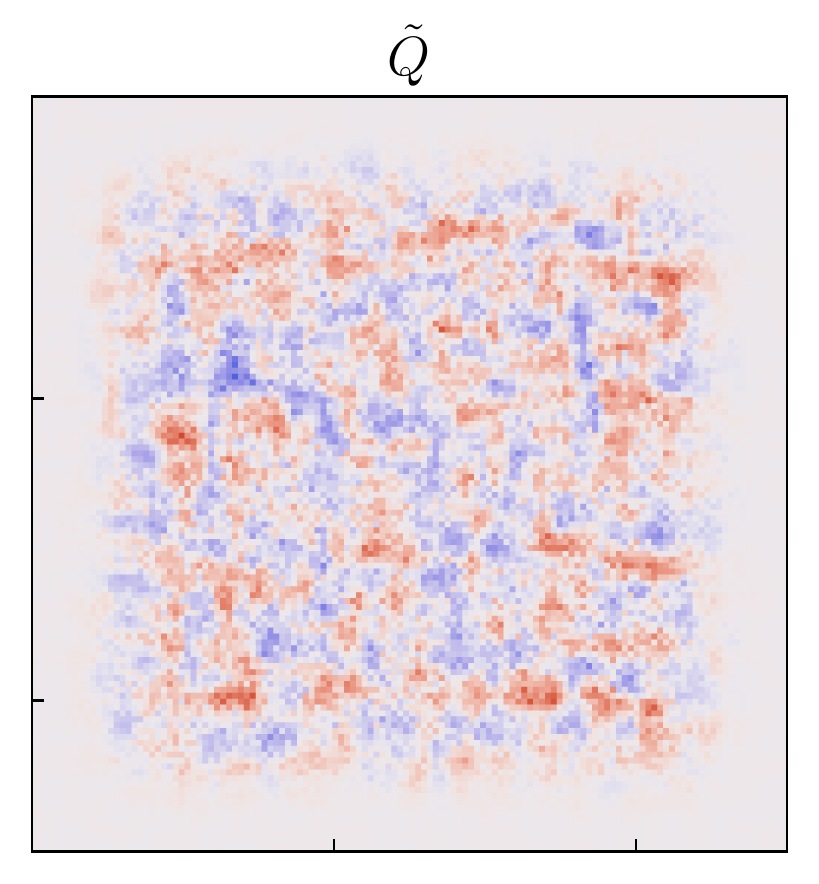}
&
\multirow{2}{*}[5.0cm]{\includegraphics[width=.1115\textwidth]{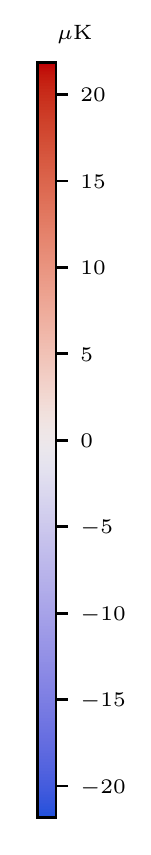}}
 \\
\includegraphics[width=.300242\textwidth]{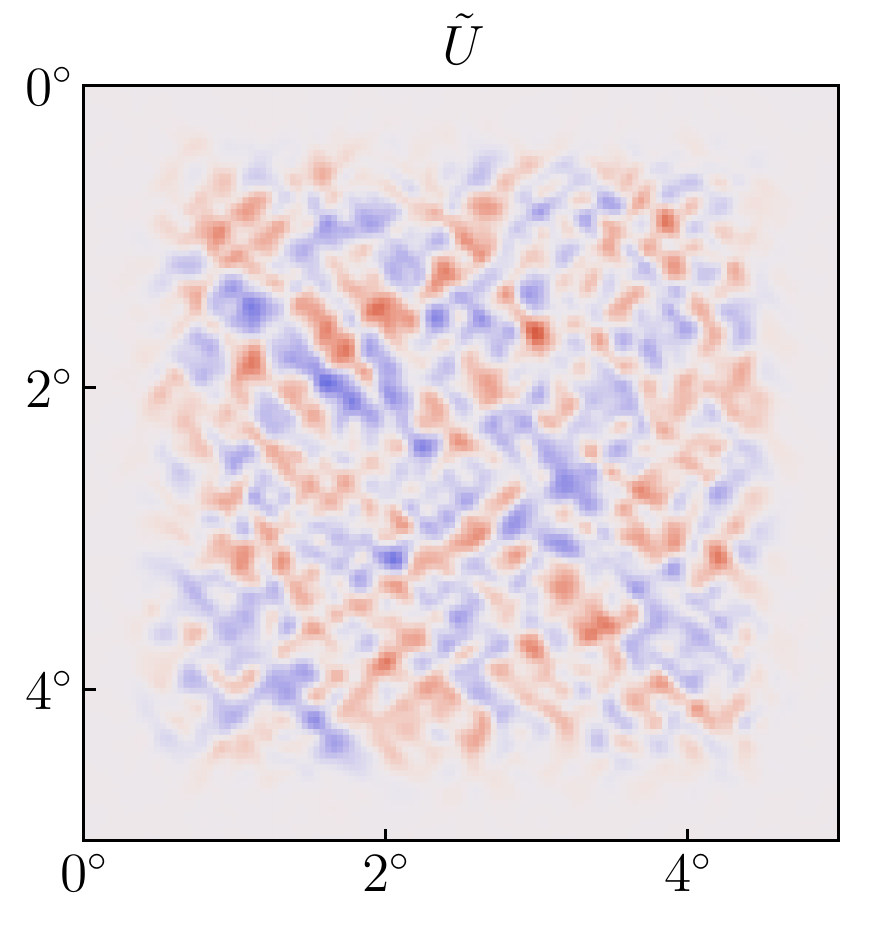}
&
\includegraphics[width=.287906\textwidth]{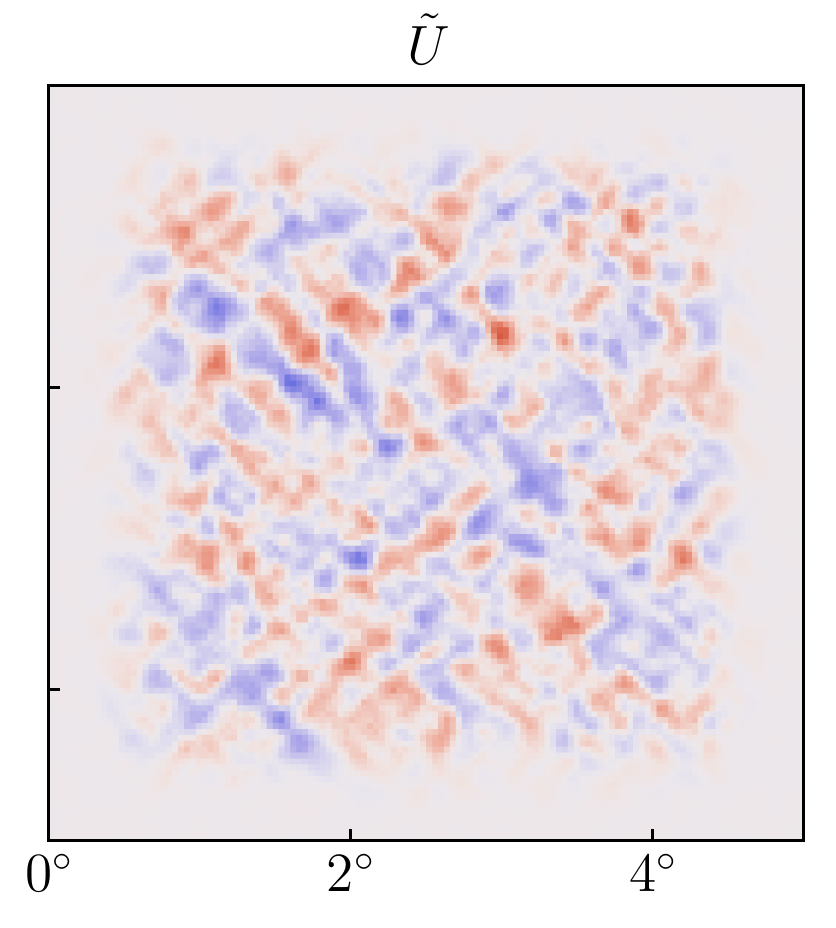}
&
\includegraphics[width=.287906\textwidth]{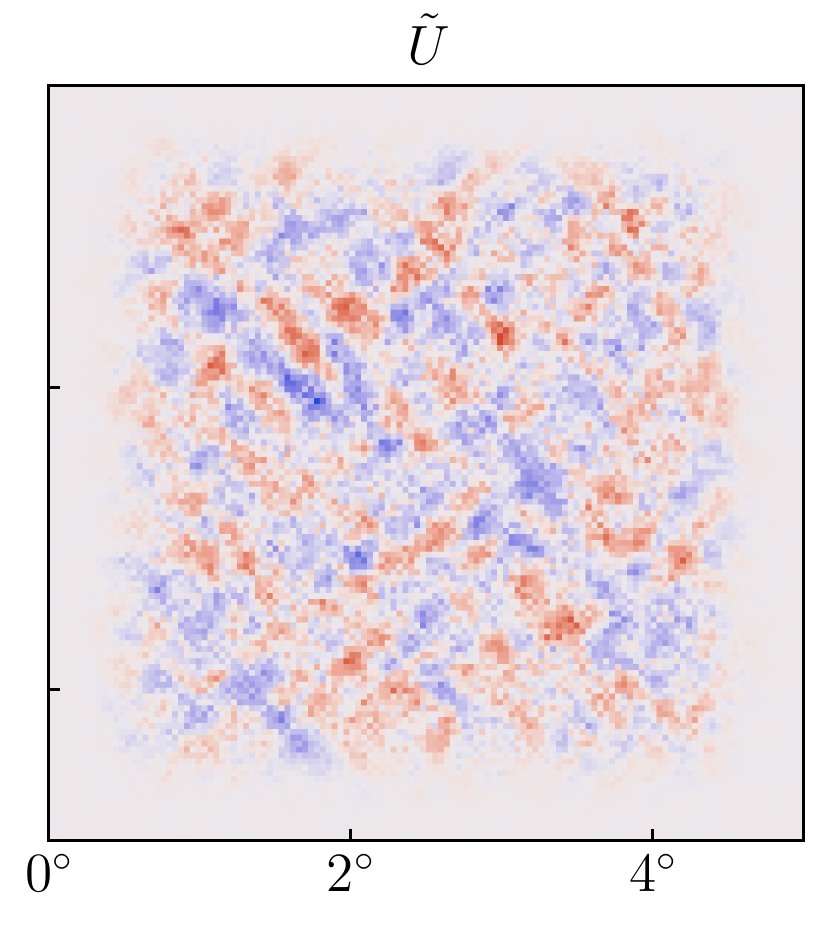}
&
\end{tabular}}
\egroup
\caption{Example of the input maps $\tilde{Q}$ (top) and $\tilde{U}$ (bottom), with apodization applied and some of the different amounts of noise used in this work (increasing left to right), for one realization of the test set. The difference between noise levels of 0 and 1 $\mu$K-arcmin is difficult to see by eye, but 5 $\mu$K-arcmin noise is clearly visible.} \label{fig:inputmaps}
\end{figure*}

}

\def \FigureEmaps {

\begin{figure*}
\centering
\bgroup
{\setlength{\tabcolsep}{0em}
\def\arraystretch{0.} 
\begin{tabular}{r r r r r}
 &
\multicolumn{1}{c}{0 $\mu$K-arcmin} & \multicolumn{1}{c}{1 $\mu$K-arcmin} & \multicolumn{1}{c}{5 $\mu$K-arcmin} & \\
\includegraphics[valign=T,width=.235904\textwidth]{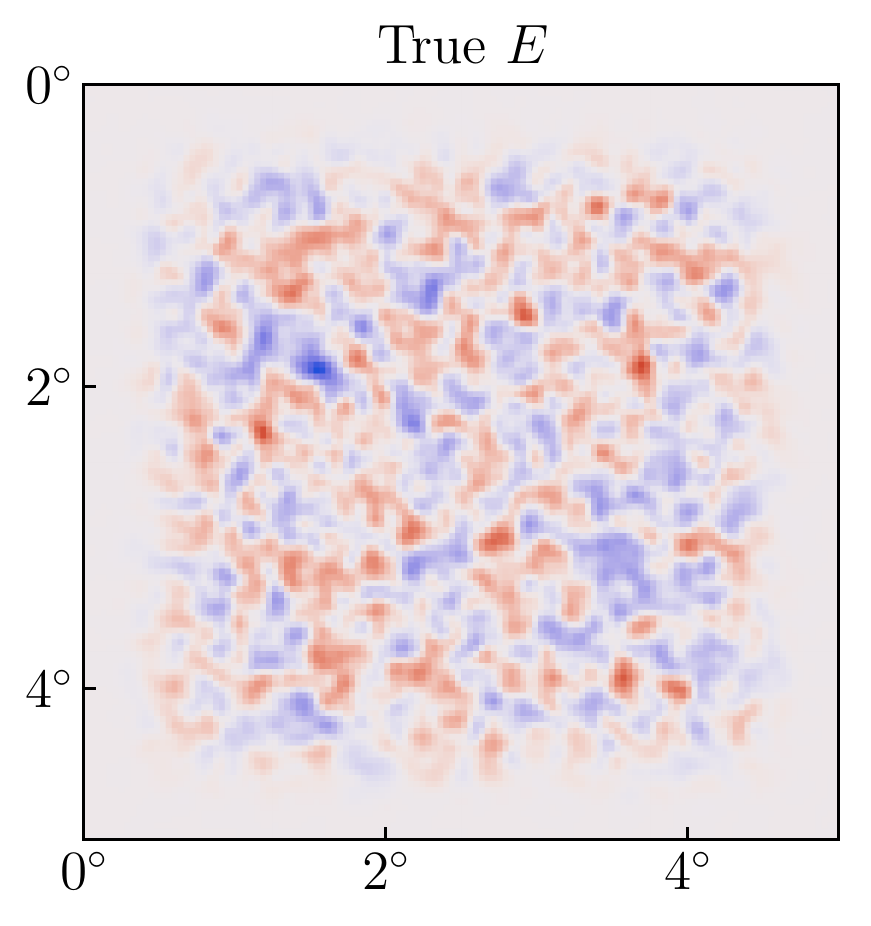}
&
\includegraphics[valign=T,width=.22\textwidth]{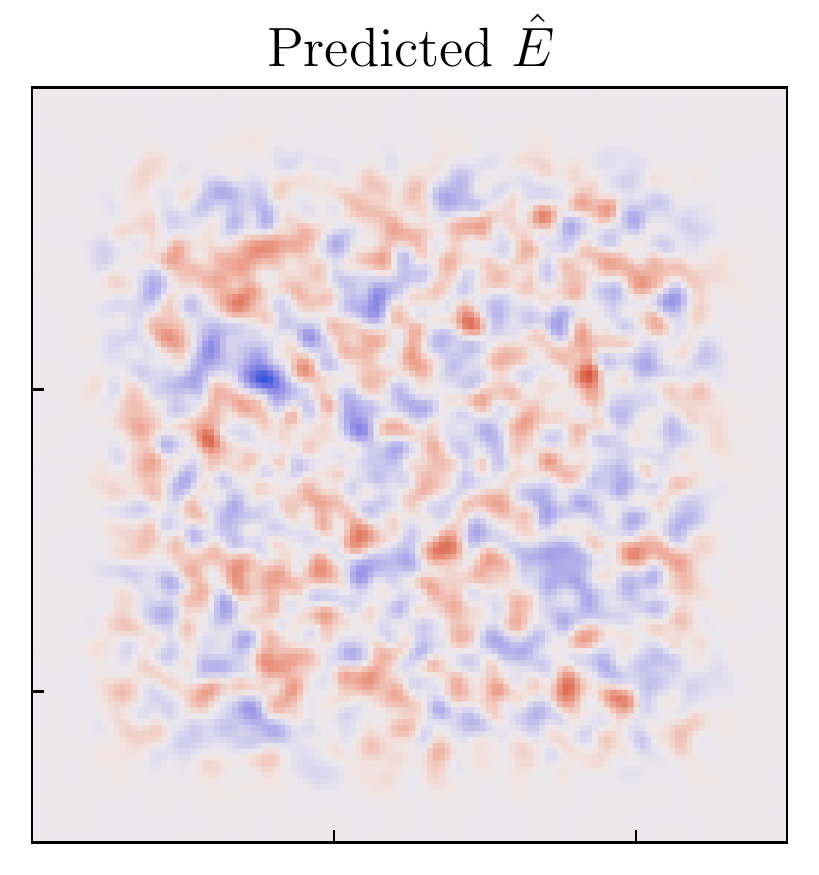}
&
\includegraphics[valign=T,width=.22\textwidth]{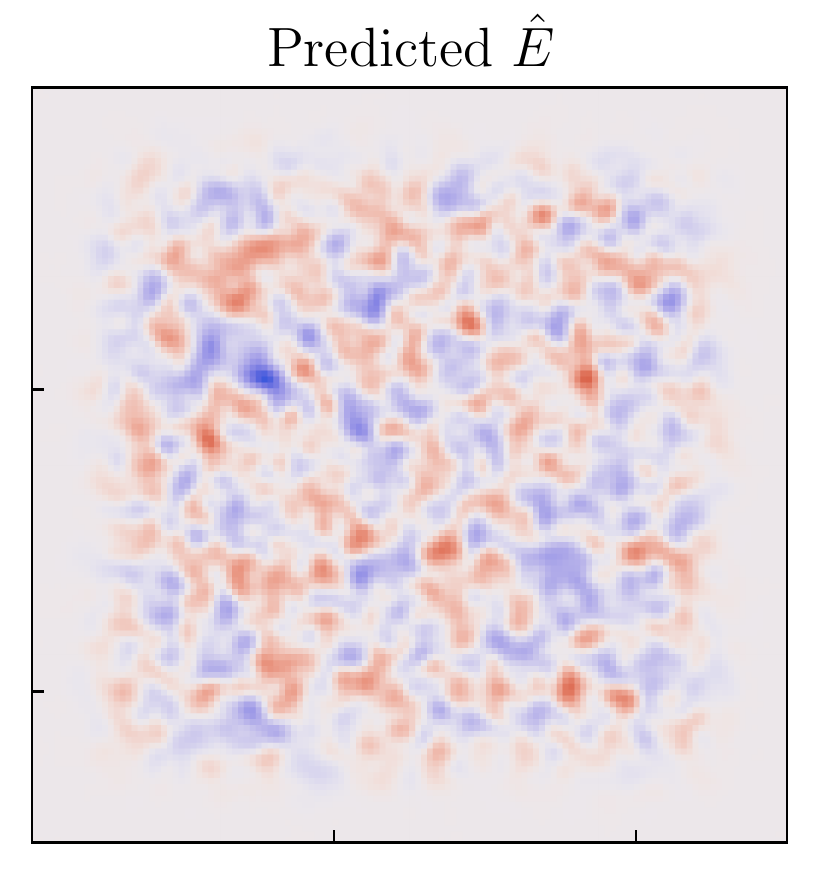}
&
\includegraphics[valign=T,width=.22\textwidth]{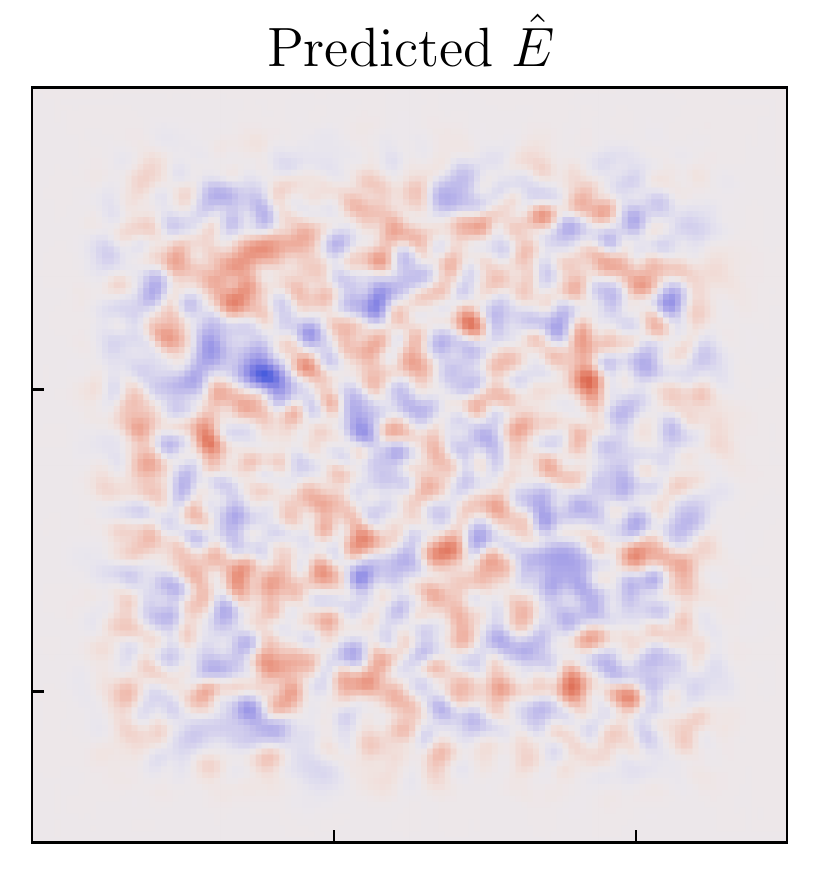}
&
\multirow{2}{*}[.24cm]{\includegraphics[width=.0955\textwidth]{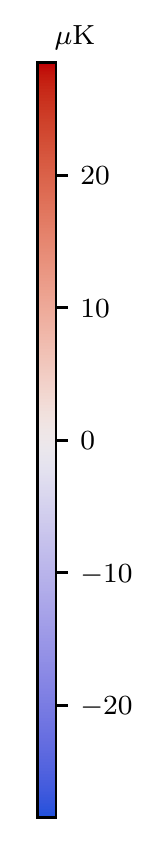}}
 \\
 &
\includegraphics[width=.235904\textwidth]{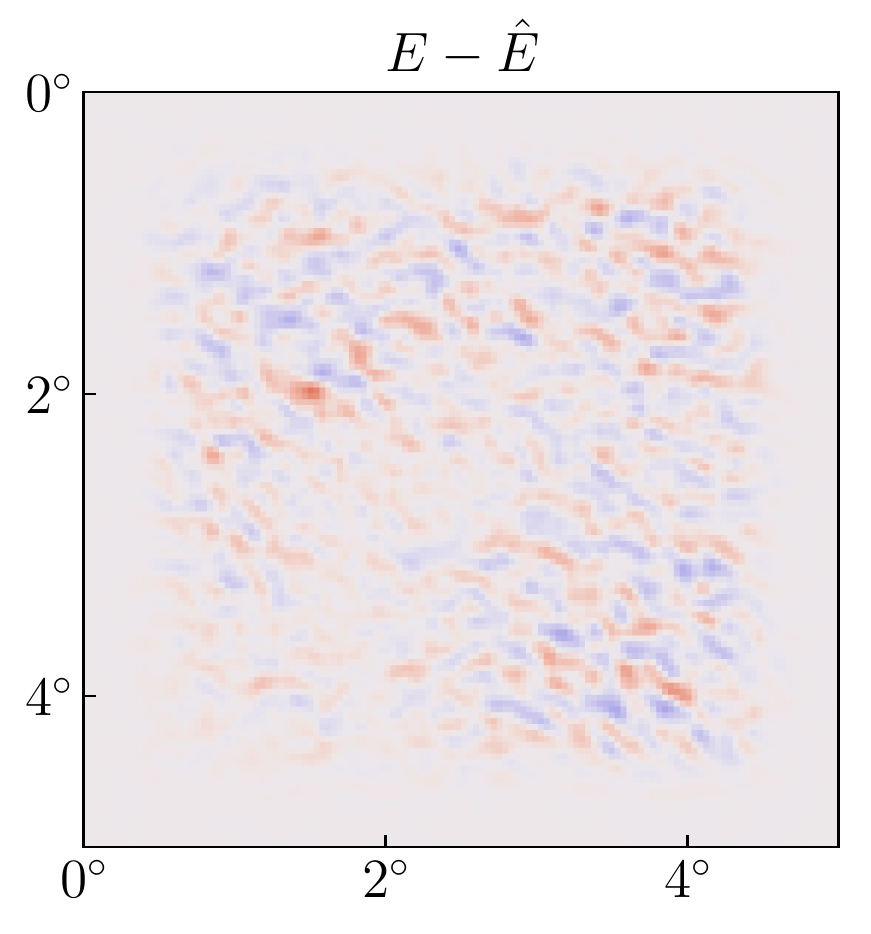}
&
\includegraphics[width=.226212\textwidth]{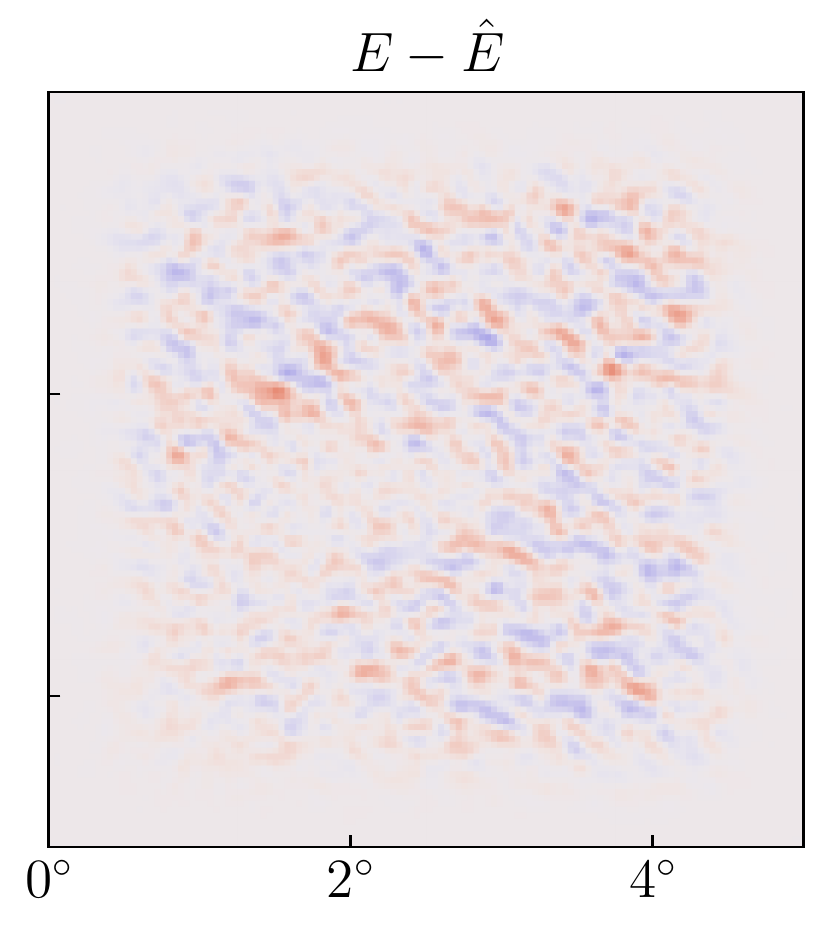}
&
\includegraphics[width=.226212\textwidth]{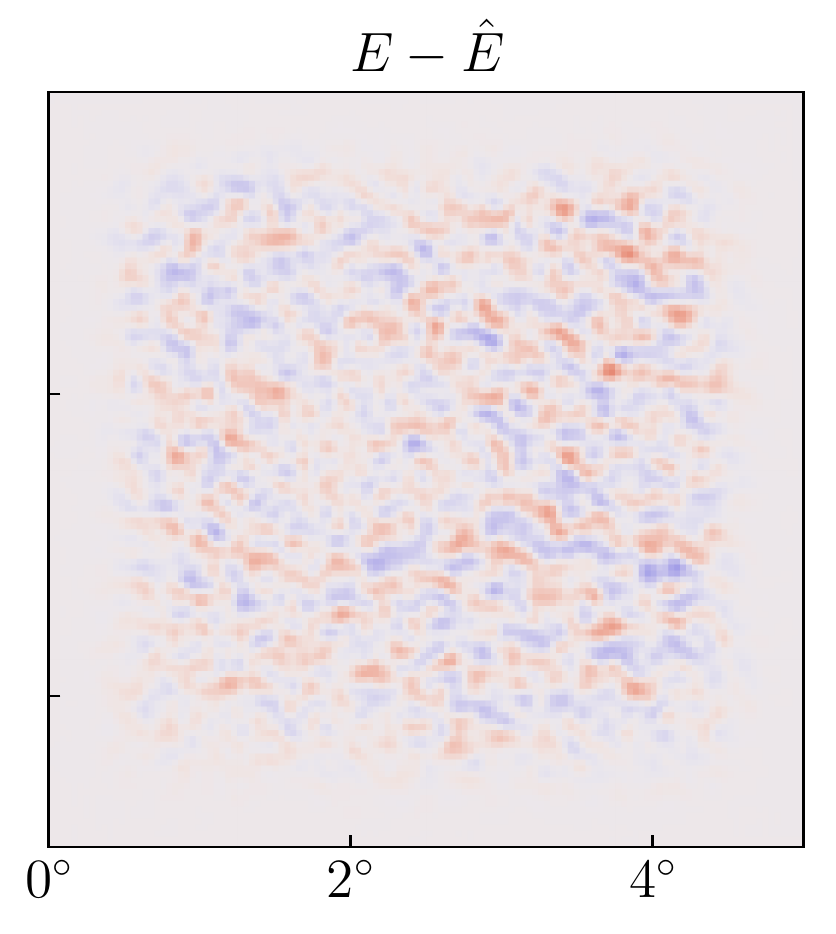}
&
\end{tabular}}
\egroup
\caption{Example of $E$-mode maps for the realization corresponding to the $(\tilde{Q},\tilde{U})$ maps shown in \myfigure~\ref{fig:inputmaps}. The true map ($E$) is shown on the left. The ResUNet predictions $\hat{E}$ (top) and the related residuals $E-\hat{E}$ (bottom) are shown for increasing levels of input noise (0, 1, 5$\mu$K-arcmin; left to right). Comparing the true and predicted maps, most of the larger-scale structure is recovered, but some visible structure remains in the residual maps. While the amplitudes of the residual maps increase with noise, the difference between the different levels of noise is not immediately visible from the predicted maps.
} \label{fig:Emaps}
\end{figure*}

}

\def \Figurekmaps {

\begin{figure*}
\centering
\bgroup
{\setlength{\tabcolsep}{0em}
\def\arraystretch{0.} 
\begin{tabular}{r r r r r}
 &
\multicolumn{1}{c}{0 $\mu$K-arcmin} & \multicolumn{1}{c}{1 $\mu$K-arcmin} & \multicolumn{1}{c}{5 $\mu$K-arcmin} & \\
\includegraphics[valign=T,width=.235904\textwidth]{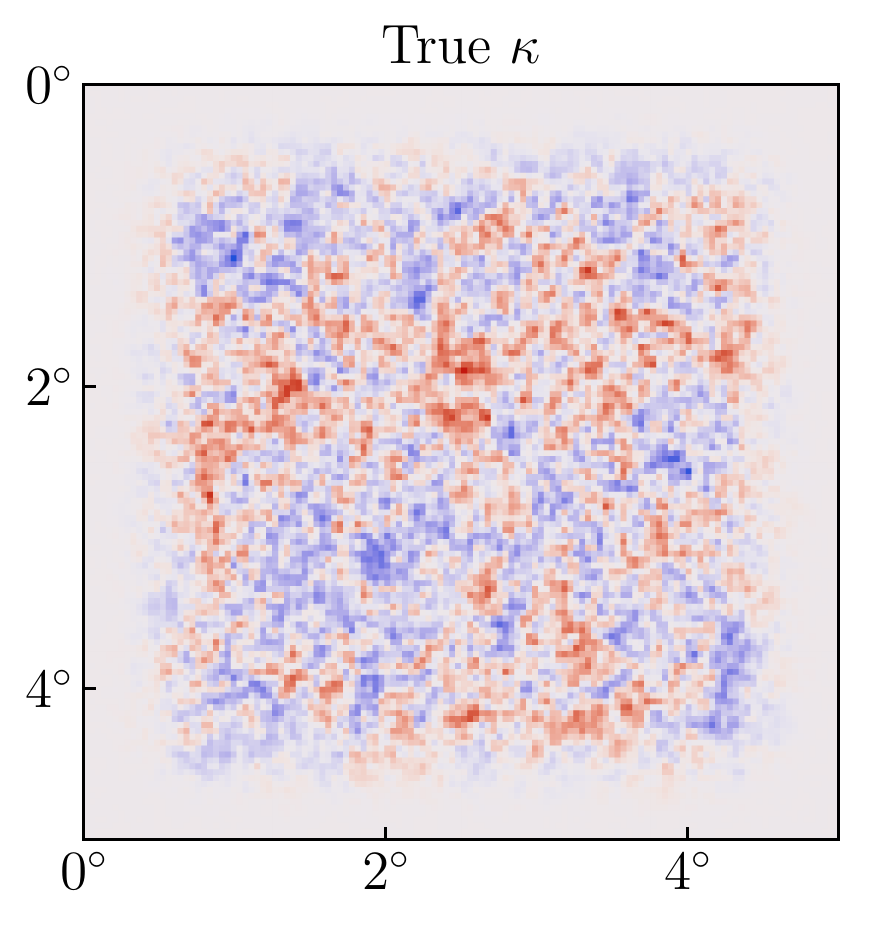}
&
\includegraphics[valign=T,width=.22\textwidth]{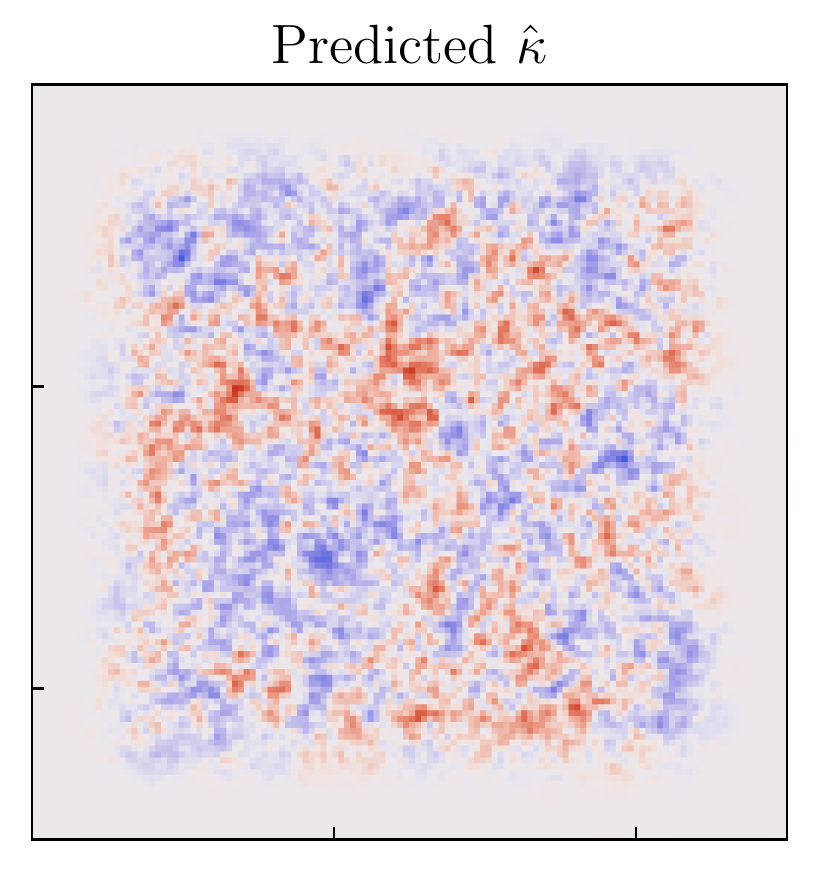}
&
\includegraphics[valign=T,width=.22\textwidth]{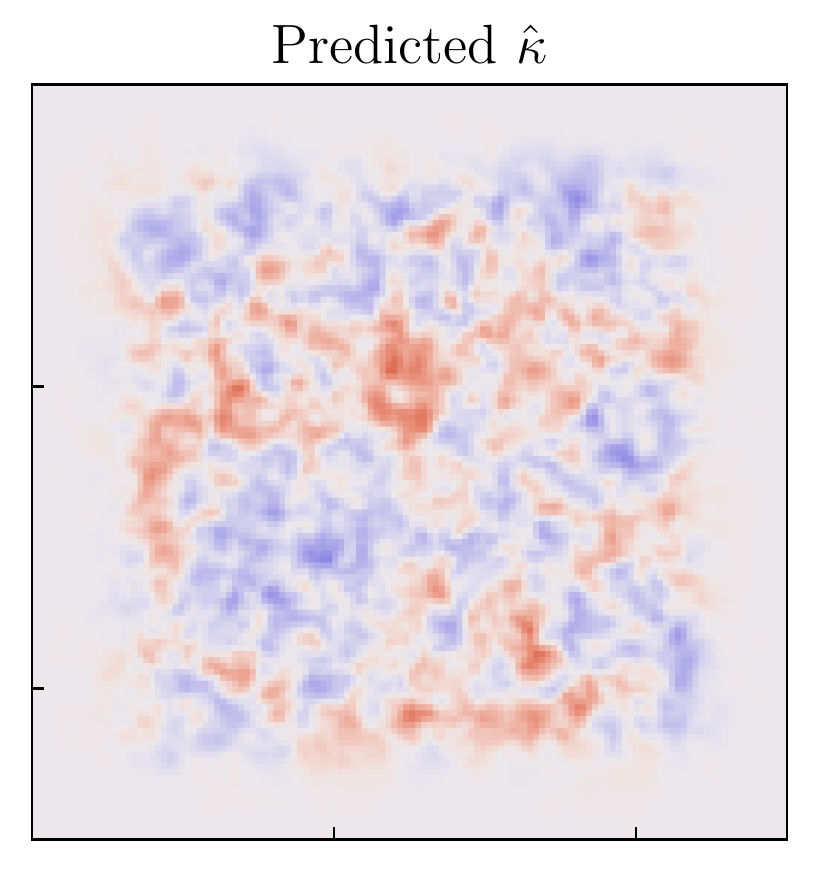}
&
\includegraphics[valign=T,width=.22\textwidth]{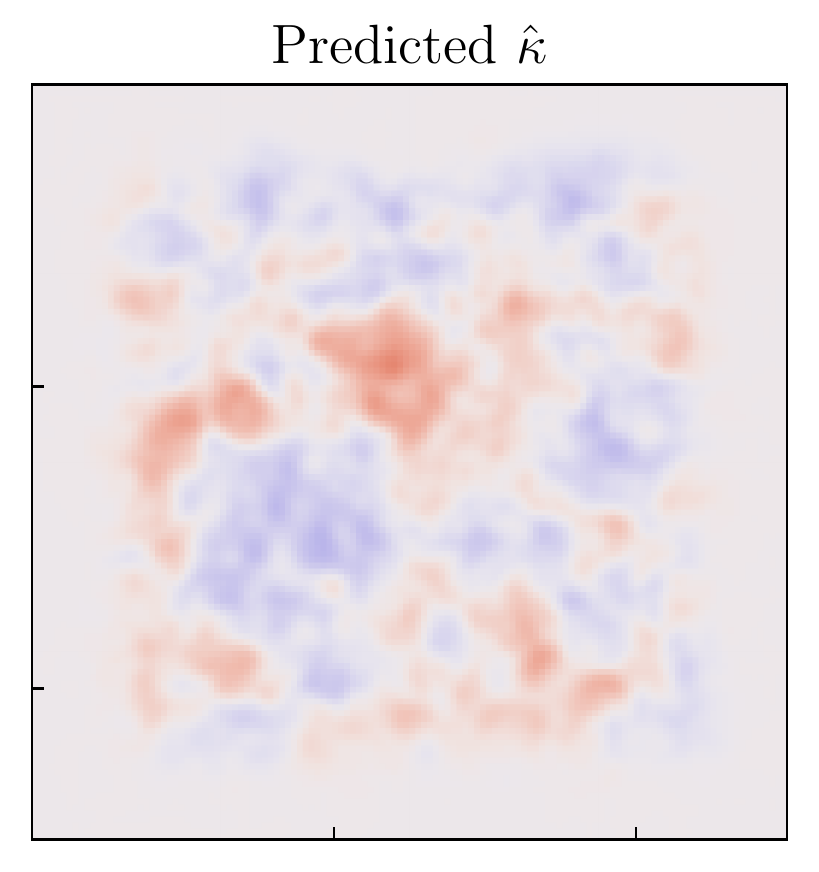}
&
\multirow{2}{*}{\includegraphics[width=.1015\textwidth]{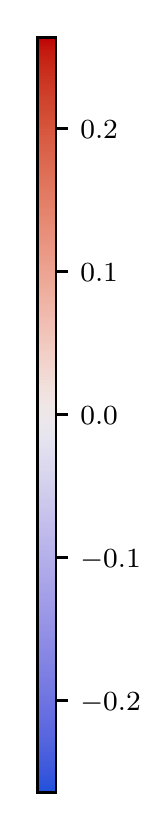}}
 \\
 &
\includegraphics[width=.235904\textwidth]{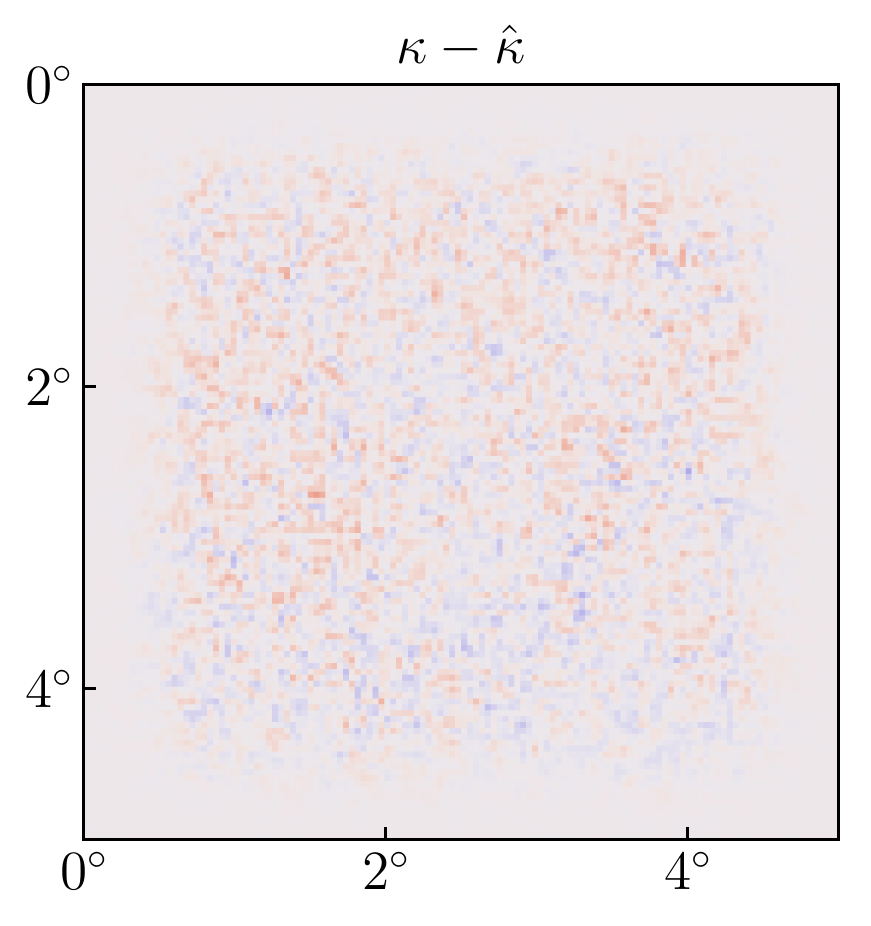}
&
\includegraphics[width=.226212\textwidth]{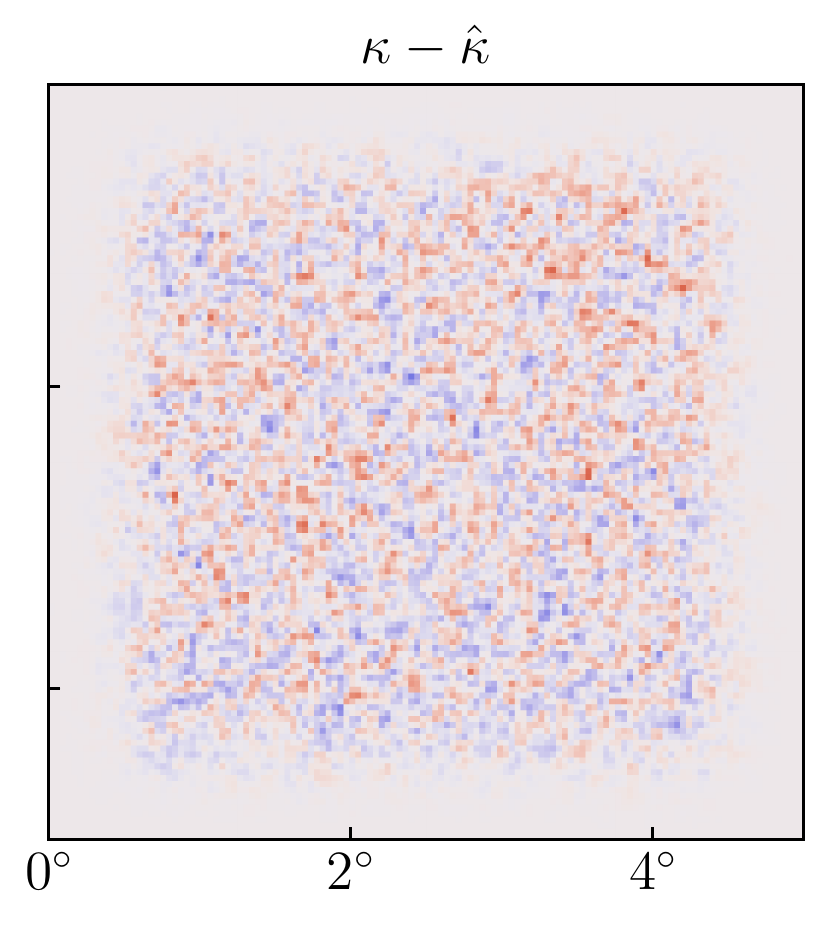}
&
\includegraphics[width=.226212\textwidth]{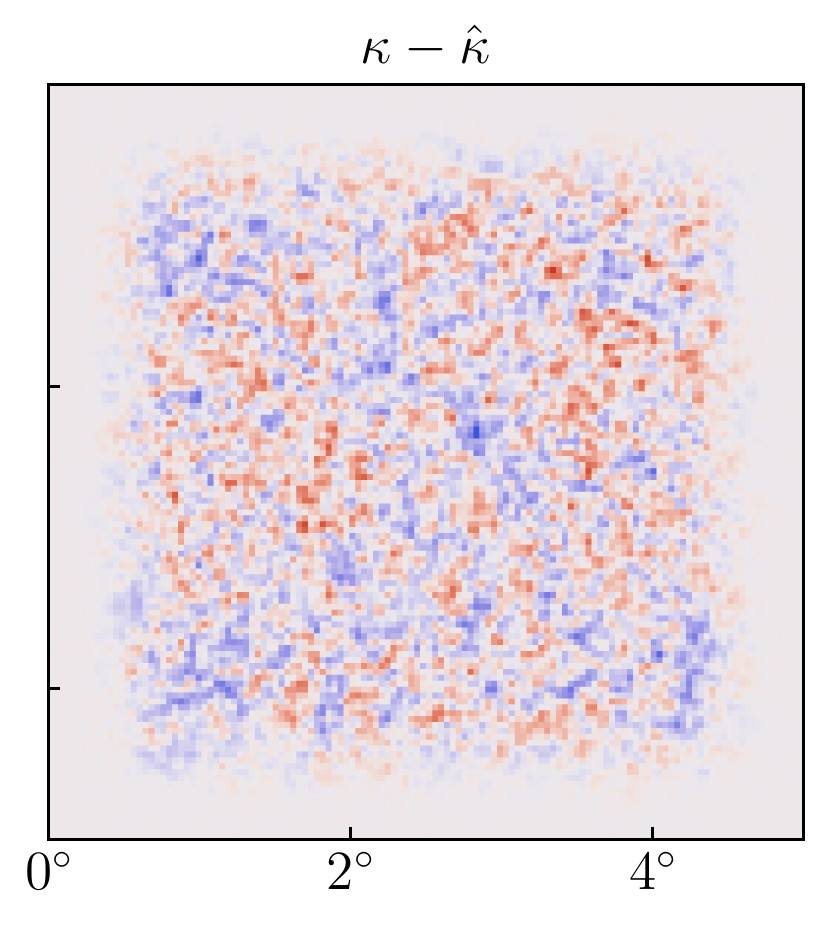}
&
\end{tabular}}
\egroup
\caption{
Example of gravitational convergence $\kappa$ maps for the realization corresponding to the ($\tilde{Q},\tilde{U}$) maps shown in \myfigure~\ref{fig:inputmaps}. The true map ($\kappa$) is shown on the left. The ResUnet predictions of $\hat{\kappa}$ (top) and the related residuals $\kappa-\hat{\kappa}$ (bottom) are shown with increasing levels of noise (0, 1, 5$\mu$K-arcmin; left to right).
Without noise, $\kappa$ recovery is better than $E$ recovery, and this is reflected here by the lack of large-scale structure in the left-most residual map. However, $\kappa$ recovery suffers much more from the addition of noise to the inputs than $E$ recovery, and once we reach 5 $\mu$K-arcmin only large-scale structure is visible in the predicted map.} \label{fig:kmaps}
\end{figure*}

}

\def \FigureFunctionqutoekv2{

\begin{figure*}
\centering
\bgroup
{\setlength{\tabcolsep}{0em}
\def\arraystretch{0.} 
\begin{tabular}{r c l}
\includegraphics[valign=T,width=.28\textwidth]{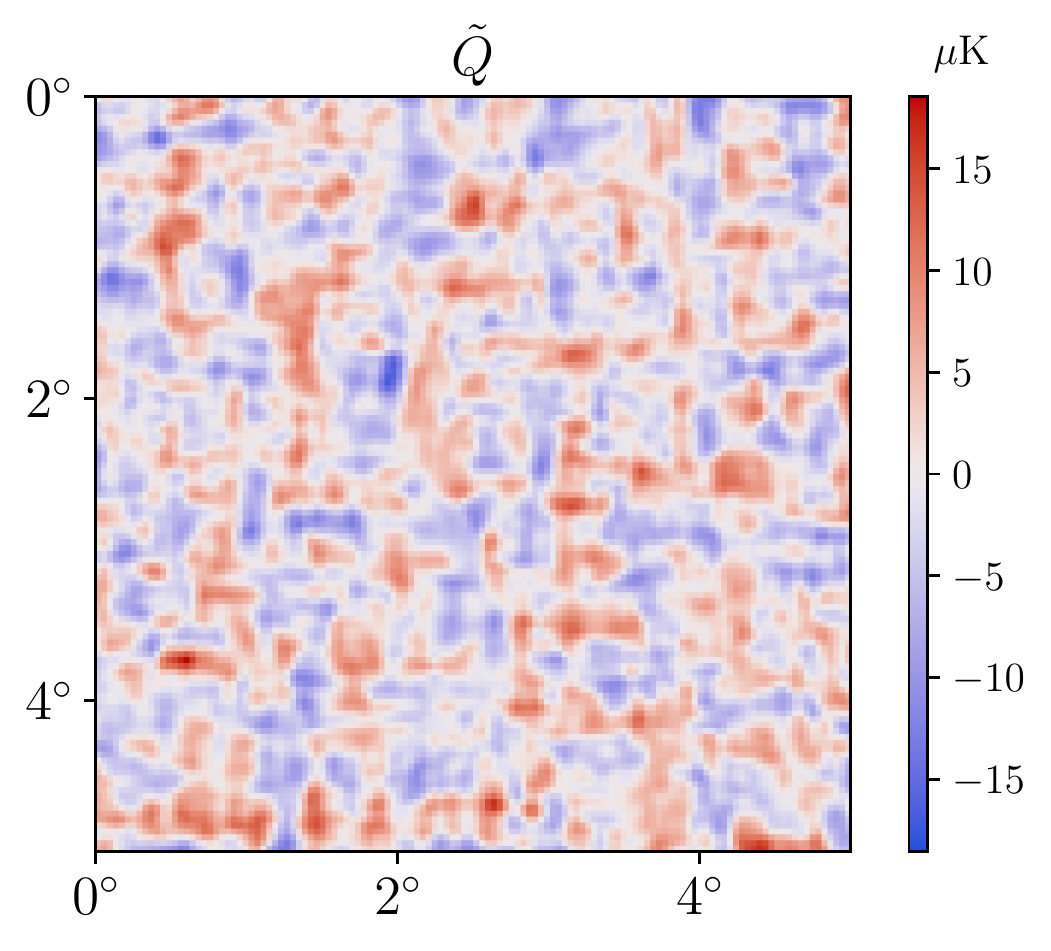}
&
\multirow{2}{*}[-.15cm]{\begin{tikzpicture}
\draw [decorate,decoration={brace,amplitude=10pt,mirror,raise=4pt},yshift=0pt]
(0,0) -- (0,8) node [black,midway,xshift=0.8cm] {};
\draw [decorate,decoration={brace,amplitude=10pt,raise=4pt},yshift=0pt]
(4,0) -- (4,8) node [black,midway,xshift=0.8cm] {};
\draw[-triangle 60] (.6,4) -- (3.4,4) node[midway,above]{ResUNet};
\end{tikzpicture}}
&
\hspace*{-1.cm} \includegraphics[valign=T,width=.28\textwidth]{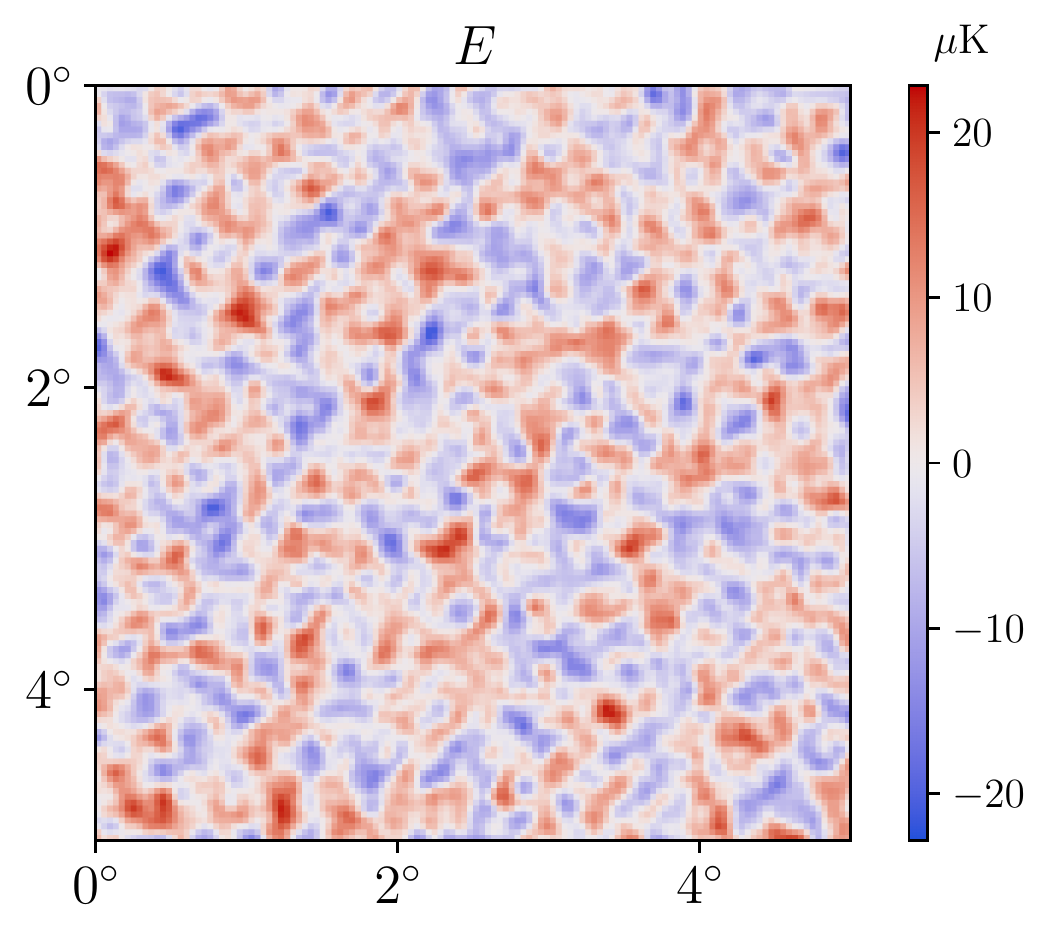} \\
\includegraphics[valign=T,width=.28\textwidth]{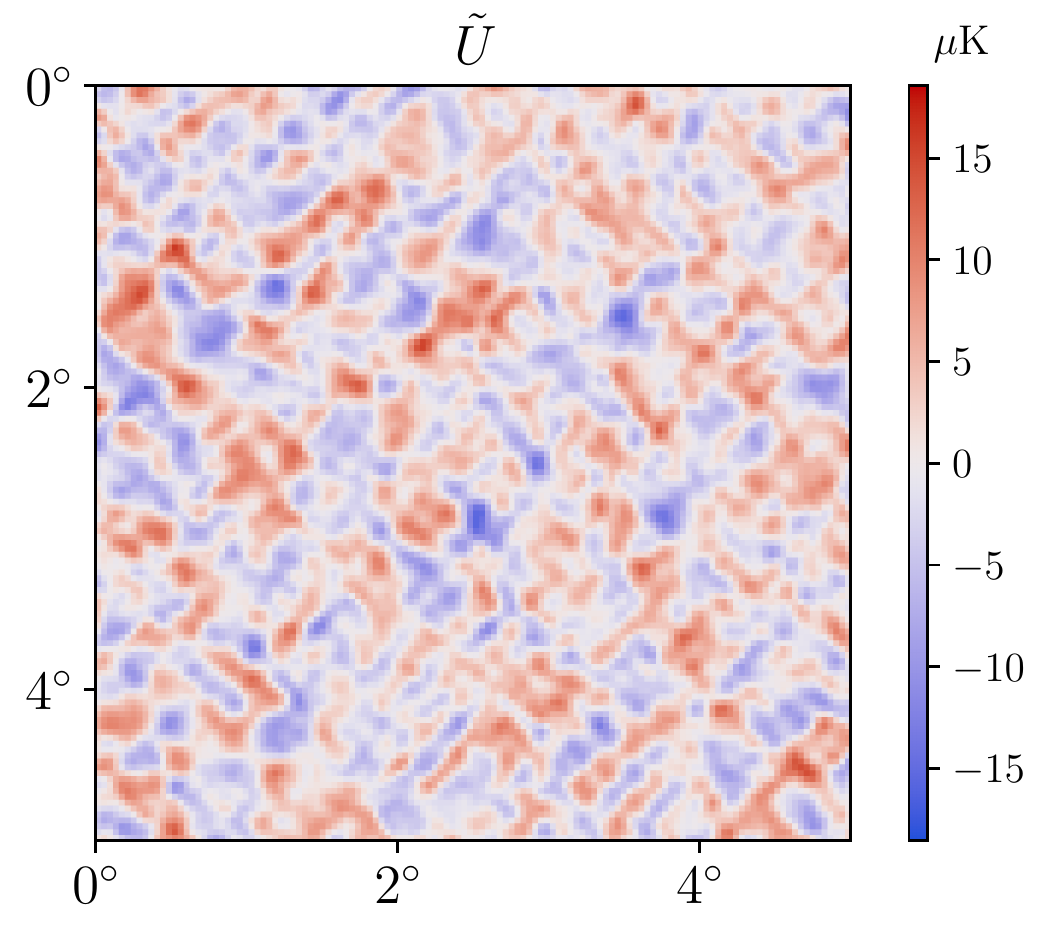}
&
&
\hspace*{-1.cm} \includegraphics[valign=T,width=.28\textwidth]{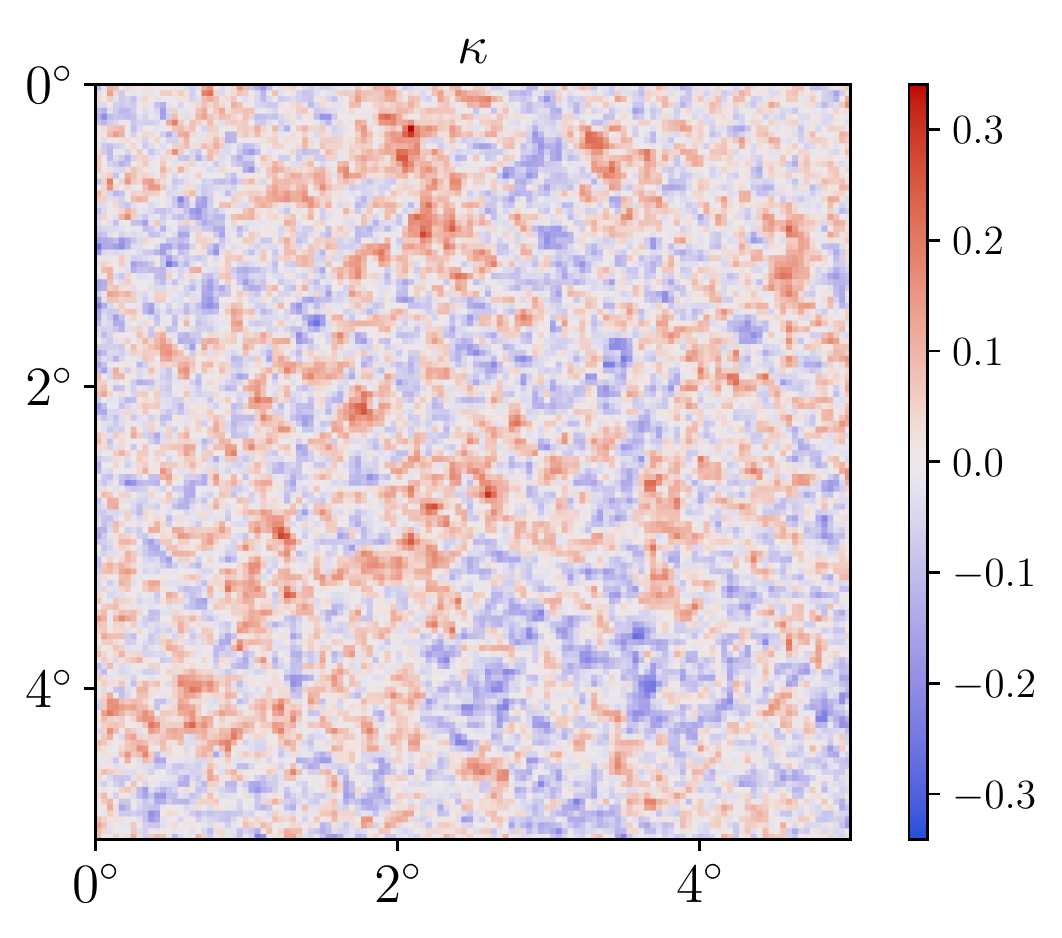}
\end{tabular}}
\egroup
\caption{We train neural networks to learn a mapping from the lensed $(\tilde{Q}, \tilde{U})$ maps into the unlensed $E$ map and the gravitational convergence map $\kappa$, extracting the underlying fields from the observed quantities. Here we illustrate this mapping using one of the realizations in the training set. The maps correspond to a patch of the sky five degrees across.} \label{fig:mapping}
\end{figure*}

}

\def \FigureNullTest {

\begin{figure}
\centering
  \includegraphics[width=\linewidth]{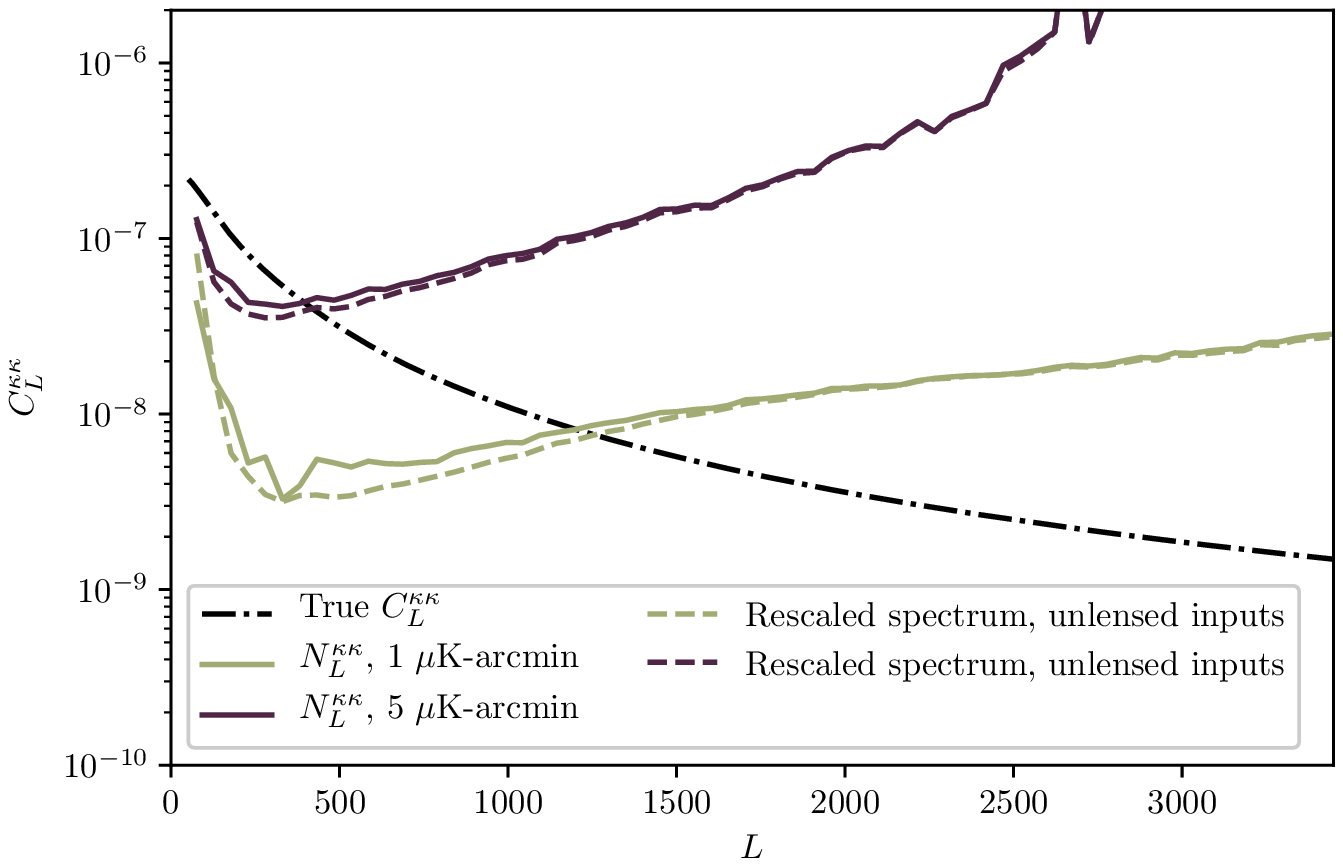}
\caption{To test how robust the network is to inputs with different levels of lensing, we look at the lensing maps predicted by the network when we pass unlensed versions of the test set inputs through it. We see that the rescaled outputs have a spectrum that is within a few percent of the noise spectra we found above. This confirms that our model of uncorrelated noise is accurate.} \label{fig:nulltest}
\end{figure}

}

\def \FigureUncertainty {

\begin{figure}
\centering
  \includegraphics[width=\linewidth]{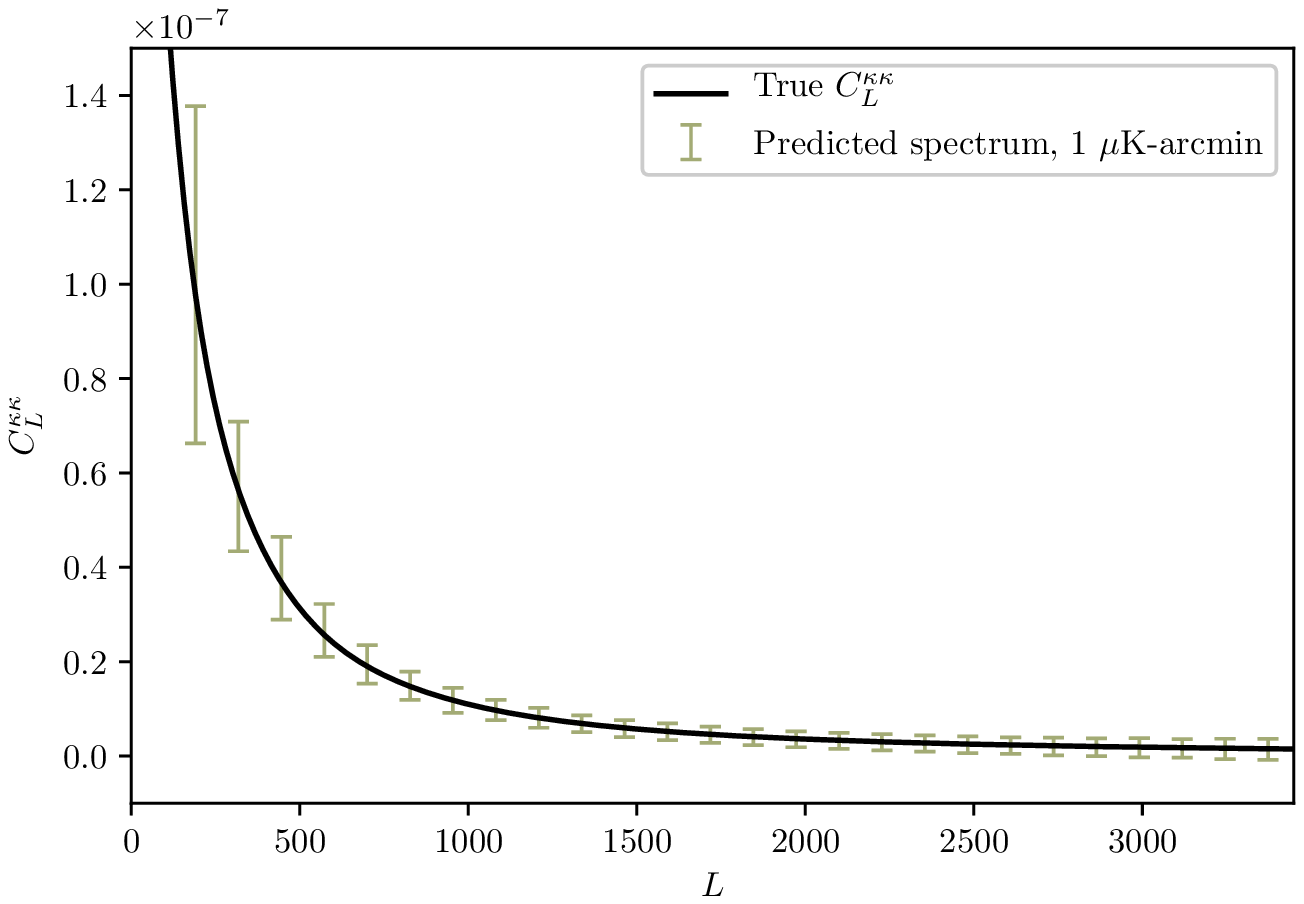}
\caption{To quantify the variance of the network output on different simulation realizations, we rescale and then remove the noise spectrum from each recovered $\hat{\kappa}$ with 1 $\mu$K-arcmin input $(\tilde{Q}, \tilde{U})$ maps. We then find the 1-$\sigma$ deviation around the mean spectrum. This variation comes both from the difference in spectrum between each simulation (cosmic variance) and from the noise in the measurements. It is the uncertainty in a measurement of the spectrum from a single simulation realization or on real data.} \label{fig:uncertainty1}
\end{figure}

}

\begin{document}

\begin{frontmatter}

\title{DeepCMB: Lensing Reconstruction of the Cosmic Microwave Background with Deep Neural Networks}

\author[efi,fnal]{J.~Caldeira\corref{cor1}}
\ead{caldeira@fnal.gov}
\author[kicp]{W.~L.~K.~Wu}
\author[kicp,uofcaa,fnal]{B.~Nord}
\author[efi,kicp]{C.~Avestruz}
\author[icerm]{S.~Trivedi}
\author[story]{K.~T.~Story}

\address[efi]{Enrico Fermi Institute \& Kadanoff Center for Theoretical Physics, University of Chicago, Chicago, IL 60637, USA}
\address[kicp]{Kavli Institute for Cosmological Physics, University of Chicago, Chicago, IL 60637, USA}
\address[uofcaa]{Department of Astronomy and Astrophysics, University of Chicago, 5640 S. Ellis Ave., Chicago, IL 60637, USA}
\address[fnal]{Fermi National Accelerator Laboratory, P.O. Box 500, Batavia, IL 60510, USA}
\address[icerm]{Institute for Computational and Experimental Research in Mathematics, Brown University, Providence, RI 02903, USA}
\address[story]{Descartes Labs, Santa Fe, NM 87501, USA}

\begin{abstract}

Next-generation cosmic microwave background (CMB) experiments will have lower noise and therefore increased sensitivity, enabling improved constraints on fundamental physics parameters such as the sum of neutrino masses and the tensor-to-scalar ratio $r$.
Achieving competitive constraints on these parameters requires high signal-to-noise extraction of the projected gravitational potential from the CMB maps.
Standard methods for reconstructing the lensing potential employ the quadratic estimator (QE). 
However, the QE is known to perform suboptimally at the low noise levels expected in upcoming experiments. 
Other methods, like maximum likelihood estimators (MLE), are under active development.
In this work, we demonstrate reconstruction of the CMB lensing potential with deep convolutional neural networks (CNN) --- i.e., a ResUNet.
The network is trained and tested on simulated data, and otherwise has no physical parametrization related to the physical processes of the CMB and gravitational lensing.
We show that, over a wide range of angular scales, ResUNets recover the input gravitational potential with a higher signal-to-noise ratio than the QE method, reaching levels comparable to analytic approximations of MLE methods.
We demonstrate that the network outputs quantifiably different lensing maps when given input CMB maps generated with different cosmologies. We also show we can use the reconstructed lensing map for cosmological parameter estimation.
This application of CNNs provides a few innovations at the intersection of cosmology and machine learning. 
First, while training and regressing on images, this application predicts a continuous-variable field rather than discrete classes. 
Second, we are able to establish uncertainty measures for the network output that are analogous to standard methods.
Beyond this first demonstration, we expect this approach to excel in capturing hard-to-model non-Gaussian astrophysical foreground and noise contributions.

\end{abstract}


\end{frontmatter}

\section{Introduction}
\label{sec:introduction}

The earliest light we can observe in the Universe is the cosmic microwave background (CMB), which was emitted $\sim$ 400,000 years after the Big Bang during a period called recombination and encodes a wealth of information about the state of the Universe at and before that time. 
The CMB is a strong probe of both the geometry and the content of the Universe, as shown through a number of experiments over the past two decades --- e.g., COBE, Boomerang, WMAP, Planck, SPT, ACT \citep{mather94, lange01, bennett13, planck2018VI, louis16, henning18}. 
In particular, measurements within the last five years have provided strong evidence for the standard cosmological $\Lambda$CDM paradigm \citep[e.g.,][]{2013ApJS..208...19H, planck2015XIIIcosmoparam, louis16, henning18}. 
Upcoming and proposed CMB experiments are designed to reach unprecedentedly low levels of map noise ($<$ few $\mu$K-arcmin)~\citep{benson14,litebird, CMBS4book, simonsobs}. 
At these noise levels, CMB Stage-4, for example, is projected to be able to constrain the tensor-to-scalar ratio $r$ to a precision of $\sigma(r) \sim 5 \times 10^{-4}$, the number of relativistic species $N_{\textnormal{eff}}$ to $\sigma(N_{\textnormal{eff}})\sim 0.03$, and the sum of neutrino masses $M_{\nu}$ to $\sigma(M_{\nu}) \sim 20$ meV~\citep{CMBS4book}. 
Tight constraints on these parameters are key to the potential discovery of primordial gravitational waves from inflation ($r$), extra degrees of freedom in the early universe ($N_{\textnormal{eff}}$), and differentiating the neutrino mass hierarchy ($M_{\nu}$).

Constraining these parameters at these levels of precision relies on high signal-to-noise ratio reconstruction of the lensing potential --- the projected weighted gravitational potential along the line-of-sight between us and the CMB.
As CMB photons travel to us, their paths get deflected by the intervening mass distributions. The lensing potential is therefore a source of information about the universe, as it is sensitive to the matter power spectrum (and therefore the sum of neutrino masses).
On the other hand, lensing of the CMB distorts the CMB at recombination and degrades our ability to constrain early universe physics that made imprints on the CMB at that time. 
As a result, reconstruction of the lensing potential and removal of the effects of lensing (delensing) from observed CMB maps are key for decoding early-universe physics.
The quadratic estimator \citep[QE;][]{hu2002} is commonly used for lensing reconstruction for the current generation of CMB experiments \citep{Story2014,planck2015XVlens,sherwin16,polarbearlens,bklens}, and is close to optimal at current noise levels.
However, when the CMB map noise is reduced to a few $\mu$K-arcmin, QE will no longer be optimal~\citep{millea17}, meaning that solutions exist with lower noise.
Therefore, maximum likelihood methods will be required in order to improve the signal-to-noise of the lensing potential reconstruction from QE~\citep{hirata2003, millea17} --- though they have yet to be demonstrated on data.

In this work, we investigate and demonstrate the usage of neural networks as an alternative for lensing reconstruction of the CMB.
The model is learned through supervision of a training set that contains the relevant physics. 
This training set consists of a set of simulated maps, including observed lensed maps, corresponding unlensed maps, and maps of the gravitational convergence (related to the lensing potential).
The observed maps are the inputs to the neural network, and the unlensed and convergence maps are the output.
To learn a function from one set of images to another suggests an architecture with two distinct steps --- one for encoding the input map information into an efficient parametrization, and one for decoding those parameters back into the output.

In neural networks, this encoder-decoder design pattern is ubiquitous, and can be used in various tasks such as learning efficient representations of the inputs~\citep{PDP1986,Elman1988,HintoSalakhutdonov2006}, machine language translation~\citep{Cho2014,Sutskever2014,Bahdanau2016}, and semantic image segmentation~\citep{Noh2015,Shelhamer2016}.
We employ an instance from a family of network architectures, ResUNets \citep{KayalibayJS17,Zhang17}, which learns a transformation between images.
While these architectures were designed with image segmentation in mind, where the desired outcome is the assignment of a discrete set of labels to the pixels, they can be adapted to image-to-image regression, where the outputs are a continuous function of the inputs.
This is an adaptation of ResUNets that is more suited for physics applications.

Standard approaches treat lensing reconstruction and delensing as separate steps.
Recent work \citep{millea17} employed maximum likelihood methods that jointly output the lensing potential and unlensed CMB maps and demonstrated the technique on simulations.
In this work, we apply ResUNets to both the lensing reconstruction and delensing problems simultaneously, similarly to~\citet{millea17}.
We will focus on characterizing the efficacy of the lensing recovery.

The paper is organized as follows. In \S\ref{sec:cmbobservationslensing}, we present a basic background for the CMB and lensing, concluding with a statement of the problem. 
We then outline traditional CMB analysis tools for lensing reconstruction in \S\ref{sec:standardanalysis}. 
In \S\ref{sec:resunets}, we describe convolutional autoencoders and ResUNets, and present the simulated data sets with which we train and test our algorithms in \S\ref{sec:data}. 
We then describe the results of the new algorithm and its comparison with standard algorithms in \S\ref{sec:results}, with a discussion of the results and their potential in \S\ref{sec:discussion}. 
We conclude and present an outlook for future work in \S\ref{sec:conclusion}.

We use the following notation conventions.
\begin{itemize}
\item $X$: unlensed field; ``true'' CMB field.
\item $\tilde{X}$: lensed field
\item $\dbhat{X}$: unbiased estimate/prediction of $X$
\item $\hat{X}$: biased estimate/prediction
\item $\langle X\rangle$: mean over population sample
\item $X^{*}$: complex conjugate
\end{itemize}

\section{The CMB and Gravitational Lensing}
\label{sec:cmbobservationslensing}

In this section, we discuss the physical underpinnings of the CMB maps that are used for the development and testing of lensing reconstruction algorithms.

Modern CMB experiments observe the temperature anisotropies and polarization of CMB photons (and any other foregrounds) in millimeter wavelengths.
Temperature anisotropies are $\sim 0.01\%$ ($\sim$ 300$\mu$K) deviations from the mean CMB temperature of $\sim$ 2.7K.
They arise from the acoustic oscillation of the photon-baryon fluid before the cosmological epoch of recombination.
This oscillation can be sourced by both density fluctuations and primordial gravitational waves.
Given the quadrupole anisotropies in the temperature, Thomson scattering of photons with free electrons during recombination causes the CMB photons to acquire a net polarization.
For reviews, see~\citet{dodelson2003, lewis2006}.

CMB polarization maps are commonly represented in two distinct bases.
The $(Q,U)$ basis corresponds to Stokes parameters and is convenient for mapping onto from the CMB instruments' polarization detector coordinates.
Alternatively, there is the $(E,B)$ basis, which is helpful for connecting the measurements to the physics of the source of polarization \citep{seljak97, kamionkowski97}.
In particular, scalar perturbations from inflation (density fluctuations) can only source the even-parity $E$-mode polarization, while tensor perturbations (gravitational waves) can source both $E$-mode and the odd-parity $B$-mode polarizations at recombination.
Polarization signals, however, are more than an order of magnitude fainter than the temperature anisotropies, and therefore have only been mapped to high signal-to-noise on small patches of sky by ground-based CMB experiments~\citep[e.g.][]{ louis16,  polarbear17, henning18, bk15}.

As the CMB photons travel from the last scattering surface to us, their paths are 
 deflected by the gradient of the gravitational potential $\phi$, an effect called gravitational lensing:
\begin{equation}
\tilde{X}_{\pm}(\hat{n}) = X_{\pm}(\hat{n} + \nabla \phi(\hat{n})),
\end{equation}
where $X$ is the unlensed field and $\tilde{X}$ denotes the lensed field \citep[e.g.,][]{hu2000} and $X_{\pm} = Q \pm i U $.
for the polarization fields. These deflections generate distorted versions of the $Q$ and $U$ maps from the surface of last scattering. 
As a result, when transformed to the $(E,B)$ basis, some $E$ modes get converted to $B$ modes.
We call these lensing $B$ modes. This means that we are therefore guaranteed to observe some $B$ modes when we observe the CMB even if there were no primordial $B$ modes at the surface of last scattering.  
These $B$ modes have been detected only in recent years \citep{bk14, keisler15, louis16, polarbear17}, and are about 10$\times$ fainter than $E$ modes. 

We can reconstruct the lensing potential from lensed CMB maps by leveraging the cross-multipole correlations that lensing introduces into the CMB maps.
With a measurement of the lensing potential in hand, we can also remove the effect of lensing from CMB maps to recover primordial signals.
Observed CMB maps from various experiments have been used to reconstruct the projected gravitational potential, whose power spectrum yields constraints on cosmological parameters \citep[e.g., $\Omega_m$, see][]{planck2015XVlens, sherwin16, omori17, simard17}.
To reach new levels of precision, high signal-to-noise reconstructions of the lensing potential play a crucial role \citep{manzotti2017}: the shape of the lensing power spectrum is sensitive to $M_{\nu}$, whereas both $r$ and $N_{\rm eff}$ require delensing for their parameter uncertainties to reach the projected levels.

For Stage 4 CMB experiments, polarization information is expected to dominate the signal-to-noise of the lensing reconstruction~\citep{CMBS4book}.
Unlike the $T$ anisotropy maps and the $E$-mode maps, primordial signal in the $B$-mode map in the standard $\Lambda$CDM cosmology would require non-zero $r$ for fitting observations.
With $r=0$, any observed $B$ modes (in the absence of foregrounds and noise) would come from lensing itself. 
Therefore, using the $E$ and $B$ maps for lensing reconstruction is extremely clean. 
In the following, we anchor our comparisons to lensing reconstruction using the EB estimator.

Gravitational lensing can also be quantified using the \emph{gravitational convergence} $\kappa$, a scalar field that physically corresponds to weighted over-densities integrated along the line-of-sight.
Throughout this text, we will represent the lensing field using either $\kappa$ or $\phi$, as one can move between the two fields using Poisson equation.
In Fourier space, using the flat-sky approximation, this is
\begin{equation}
\kappa(\bl) = -\frac{1}{2} \ell^2 \phi(\bl),
\end{equation}
where $\bl$ is the two-dimensional vector of multipole moments. 

We defer investigations of impacts from galactic and extragalactic foregrounds to lensing reconstruction and delensing to later work. 
Therefore, all the simulations involved only contain information from the CMB maps (both lensed and unlensed) and $\kappa$. 
We set unlensed $B=0$, 
as our focus is on lensing reconstruction and primordial $B$-modes have not yet been discovered.
We do add basic realism by adding noise, beam, and apodization mask, which we describe in more detail in \S\ref{sec:data}.
The task we set ourselves is to recover the unlensed $E$ map as well as the lensing convergence map $\kappa$ from the lensed $(\tilde{Q}, \tilde{U})$ maps.
This can be treated as an image-to-image regression problem and is summarized in \myfigure~\ref{fig:mapping}.

\FigureFunctionqutoekv2

\section{Standard CMB Lensing Reconstruction Methods}
\label{sec:standardanalysis}

We will compare the neural network approach to current standards in analysis --- the quadratic estimator, and more futuristic iterative methods.
We quantify the efficiency of the neural network approach by comparing the algorithms in terms of a noise proxy, defined in \S\ref{sec:quadraticestimator}.
In the following we briefly describe the QE lensing reconstruction scheme, define the noise that we use to compare different methods, and outline the maximum likelihood noise estimate.

\subsection{Quadratic Estimator (QE)}
\label{sec:quadraticestimator}

The primordial CMB is well-approximated as Gaussian random fields and therefore can be completely described by 2-point statistics (e.g. the power spectrum, denoted by the multipole moments $C_{\ell}$).
Lensing of the CMB by the intervening gravitational potential introduces correlations between angular scales corresponding to the size of the lenses. 
Therefore, the covariance of the CMB map is no longer diagonal in $\ell$, as it would have been if it were not lensed. 

The QE method for lensing reconstruction uses the off-diagonal covariance of the lensed CMB maps to estimate the lensing potential $\phi$.
Specifically, the covariance is proportional to $\phi$: 
\begin{equation}
\left <  X(\bl) X'(\bL-\bl) \right > \propto \phi(\bL),
\end{equation} 
where $X, X'$ denote lensed CMB fields and $\langle \rangle$ denotes the average over CMB realizations. This relationship can be written down explicitly for pairs of CMB fields $(T,E,B)$ \citep[e.g.][]{hu2002} and can therefore be used to construct estimators that extract $\phi$ from the lensed fields. Indeed, the covariance would be zero if $\phi$ = 0.
In our case, the maps are the observed (lensed) $\tilde{E}$- and $\tilde{B}$-mode maps, which are converted from the Stokes $(\tilde{Q},\tilde{U})$ space. 
For further reference, the equations in this section are based on \citep{hu2002} for the EB estimator.

An unnormalized $\phi$ map in Fourier space, $\hat{\phi}$, can be estimated as
\begin{equation}
\hat{\phi}_{\bL} = \int \frac{d^2 \bl}{(2\pi)^2} w^{\phi}_{\bL, \bl} \bar{E}_{\bl} \bar{B}_{\bL-\bl}
\end{equation}
where $\bL$ and $\bl$ are two-dimensional multipole vectors, and ($\bar{E}$, $\bar{B}$) are Wiener-filtered $(\tilde{E},\tilde{B})$ maps; the filter is defined as 
\begin{equation}
\bar{X} \equiv \frac{\tilde{X}}{\left(\tilde{C}^{XX}_{\ell}+N^{XX}_{\ell}\right)},
\end{equation} 
where $N_{\ell}^{XX}$ is the noise power spectrum for the field $X$. The weight $w$ is given by
\begin{equation}
w^{\phi}_{\bL, \bl} = -\bl \cdot (\bL - \bl)\, \tilde{C}^{EE}_{\ell}\, \sin (2\psi),
\end{equation}
where $\psi$ is the angle between $\bl$ and $\bL-\bl$ \citep{hu2002}.

To obtain an unbiased estimate of $\phi$, $\dbhat{\phi}$, we subtract the mean-field that arises from masking and normalize it by $1/R$, where $R$ is the response:
\begin{align}
\dbhat{\phi}_{\bL} &= \frac{1}{R} \left( \hat{\phi}_{\bL} - \langle  \hat{\phi}_{\bL} \rangle \right) \\
\frac{1}{R} &= \int \frac{d^2 \bl}{(2\pi)^2} \frac{ |w^{\phi}_{\bL, \bl} |^2 }{ (\tilde{C}^{EE}_{\ell}+N^{EE}_{\ell}) (\tilde{C}^{BB}_{\ell}+N^{BB}_{\ell})} 
\label{eqn:qenorm}
\end{align}
Generically assuming $\hat{\phi} =  R \phi + n_{\phi}$, where one can think of $R$ as a multiplicative bias and $n_{\phi}$ as an additive bias, we can write the power spectrum of $\dbhat{\phi}$ as follows:
\begin{equation}
 \langle \dbhat{\phi}^*(\bL) \dbhat{\phi}(\bL') \rangle =   \delta(\bL-\bL')( C_L^{\phi\phi}  + N^{\phi\phi}_L ),
\end{equation}
where $C_L^{\phi\phi}$ is the input $\phi$ field's power spectrum and $N^{\phi\phi}_L$ is the noise spectrum. 
$N^{\phi\phi}_L$ is the noise term that we compare between the various algorithms.

The noise spectrum provides a numerical point of comparison, because it is informative for parameter estimation. For example, when constraining the sum of neutrino masses, the noise spectrum directly enters the Fisher forecast of 1-$\sigma$ uncertainty~\citep[e.g.][]{wu2014}.
Additionally, in the limit of noise reachable by CMB experiments in the next decade, the delensing efficiency is mostly a function of cross-correlation of the
$\dbhat{\phi}$ to the true underlying $\phi$, 
$\rho_L = C^{\dbhatss{\phi}\phi}_L / \sqrt{ (C^{\phi\phi}_L + N^{\phi\phi}_L) C^{\phi\phi}_L }$,
which is also directly related to the noise spectrum. 

In this work, we extract $N^{\phi\phi}_L$ from the QE by 
\begin{equation} \label{eqn:phinoisedefn}
N^{\phi\phi}_L = \langle C^{\dbhatss{\phi} \dbhatss{\phi}}_L \rangle - \langle C^{\phi \phi}_{L} \rangle,
\end{equation}
the difference between the ensemble average of the auto-spectra of the 
estimated $\phi$ and that of the input $\phi$. 
$N^{\phi\phi}_L$ is typically estimated from simulations in standard analysis, calculated as $N_0$ and $N_1$ noise biases.
$N_0$ denotes the disconnected 4-point term that is 0th order in $C^{\phi\phi}_L$, while $N_1$ is 1st order in $C^{\phi\phi}_L$~\citep{kesden03}. $N_0$ is the largest contribution to $N^{\phi\phi}_L$. 
Here, since we are working with simulations, instead of calculating $N_0$ and $N_1$ directly, the difference as defined in \myequation~\eqref{eqn:phinoisedefn} suffices for comparison.
It would capture $N_0$, $N_1$ and any other noise source that does not correlate with the input.

\subsection{Iterative Estimator}
\label{sec:maximumlikelihood}
We use the algorithm presented in~\citet{smith10} to estimate the noise achievable by maximum likelihood estimators. 
The idea is based on iterating the quadratic estimator for $\phi$ with CMB maps that are delensed with the estimated $\phi$.
Supposing that the lensing $B$ modes are the only $B$-mode contribution, one can reconstruct $\phi$ from $E$- and $B$-mode maps, and the $\phi$ map can be used to remove the lensing $B$ modes in the input map. 
One can then use the delensed $B$-mode map in combination with the $E$-mode map to estimate the remaining $\phi$ field, and then use this estimated $\phi$ to delens the $B$-mode map. 
These two steps can be iterated until the residual $B$ modes no longer get reduced.

It was found that the $N_0$ noise computed in this approach asymptotes towards maximum likelihood estimators, as presented in~\citet{hirata2003}.
We therefore compare the noise proxy from our neural network approach to this $N_0$ to get a sense of how closely the neural network estimator gets to maximum likelihood $\phi$ estimators. 
This is a reasonable comparison because the analytic estimate provides a theoretical lower limit on the noise of the reconstructed lensing potential power spectrum.
If the neural network recovered $\phi$'s noise spectrum approaches this, it would provide an argument for the utility of this estimator for next-generation CMB experiments.

Note that comparing the analytic $N_0$ from the iterative estimator to the noise spectrum from the output of the network is not strictly an apples-to-apples comparison. In particular, we expect the analytic $N_0$ to perform better than $N_0$ extracted from realistic simulations, as the analytic $N_0$ does not capture effects like masking. On the other hand, the comparison between the QE noise spectrum and the network output noise spectrum is an apples-to-apples comparison.

\section{Deep Learning Reconstruction Method: Residual U-Nets (ResUNet)}
\label{sec:resunets}

A growing number of physics tasks utilize machine learning techniques (see \citet{Mehta18} for a recent review). Our task is to create a network that can learn the mapping from the observed lensed $(\tilde{Q},\tilde{U})$ images to unlensed CMB images and the gravitational convergence map $\kappa$.  In this application, we use a type of \emph{feed-forward deep neural network} called a Residual U-Net (ResUNet) \citep{KayalibayJS17,Zhang17}.

The fundamental building block of a feed-forward neural network is a neuron, which receives (typically scalar) inputs, and outputs a real number.
Much of the power of neural networks stems from how neurons are connected to each other. A neural network is organized in a number of layers, each layer comprising a set of neurons.
The first layer's inputs are the inputs of the network, while the final layer gives the network output.
For instance, in our problem the output layers have a total of $2 \times 128 \times 128$ neurons, each corresponding to a pixel of one output map.
\emph{Deep} neural networks typically refer to neural networks having more than (usually much more than) 3 to 4 layers.

Information is propagated forward layer-by-layer through the network from the inputs to the outputs, hence the terminology \emph{feed-forward}.
Every neuron takes a linear combination of the outputs of a subset of neurons in the previous layer, and then applies a non-linear function, known as the \emph{activation function}, to that combination.
The composition of these simple operations from the first to the final layer can result in a highly non-linear mapping between the input and the output images.
The multiple weights in each of the linear combinations are optimized using gradient descent applied to the error in the output.
As the gradient uses the chain rule to move back from the outputs to each layer, this process is called \emph{backpropagation}.

When working with images, the most ubiquitous type of neural network is a \emph{Convolutional Neural Network} (CNN),
which is defined as a feed-forward neural network with at least one convolutional layer, named so because it implements a discrete convolution.
Each neuron in a convolutional layer takes input from  neurons in the previous layer located inside a $n \times n$ window centered at its position.
Typical values for $n$ range within $n = 3,5,7$, and the transformation performed by the convolutional layer on the window is a \emph{filter}.
Crucially, the network keeps weights of the linear combinations independent of the position in the image.
This parameter sharing reduces the complexity of the network and explicitly encodes translational equivariance.

The output size of a convolutional layer is controlled by three parameters: Number of convolutional filters applied, \emph{stride}, and amount of zero padding.
The stride may be defined as the distance in pixels between the centers of adjacent filters.
With appropriate zero padding around the image, and sliding the convolution filter with a stride of 1, the output map will be of the same size as the input.
Likewise, we can reduce the size of the output map into half by choosing a stride of 2.
The size of the output map can also be doubled by up-sampling the input, i.e., introducing zeros between pixels of the input.
A review of convolutional layers and their arithmetic can be found in \citet{Dumoulin2016}.

Convolutional layers are a natural way to take spatial context into account, as each pixel is only a function of the pixels in the previous layer that are contained inside the window defined by the convolutional filter.
As we stack convolutional layers on top of each other, the region of the input that any given pixel is a function of increases.
The size of this region at any specific layer is called \emph{receptive field} of the layer.
The fact that the receptive field increases as we move from layer to layer makes it so that each layer is sensitive to features at increasingly larger scales (corresponding to lower $\ell$ modes), allowing for both local and global information to propagate through the network.

We choose the architecture of our neural network such that the network first encodes relevant information from the input maps into smaller maps, and then decodes that information to form the output maps.
The canonical example of this design pattern is an autoencoder \citep{PDP1986,HintoSalakhutdonov2006,Elman1988}, for which the desired outputs are equal to the inputs, 
and which have applications in dimensionality reduction, compression, and unsupervised feature learning. 
More generally, the encoder-decoder strategy can be employed to learn compact representations of mappings that are not necessarily the identity function:
The encoder part of the network learns the important features of the input at different scales, and the decoder combines these features into more and more complicated representations.
This is our goal in this work. 

To achieve this, we implement a UNet \citep{RonnebergerFB15}, which takes the simple encoder-decoder with convolutional layers, and adds extra shortcuts (\emph{skip connections}) between the encoding and decoding layers to allow for propagation of small-scale information that might be lost when the size of the images decreases.
UNets were first introduced as a method for image segmentation in a biomedical context \citep{RonnebergerFB15}, and have a recent application in physics, where they were used to process sea surface temperature measurements and aid in the prediction of future sea surface temperature \citep{deBezenac17}.
Notably, physics applications of these architectures, such as in \citet{deBezenac17} and this work, use UNets for a regression task with a continuous output variable.
This is distinct from the typical image semantic segmentation, where the outputs are discrete labels applied to each pixel.

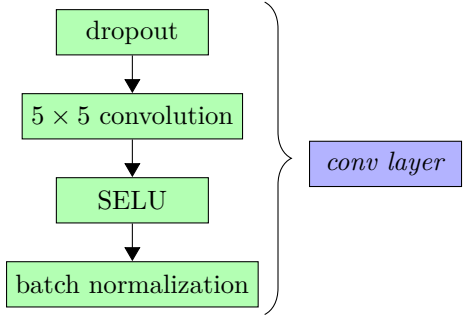
\begin{figure}[t]
\begin{center}
  \begin{tikzpicture}
\node[draw, minimum height = .6cm, minimum width = 2cm,fill=green!30] (state0){dropout};
\node[draw,below=.5cm of state0, minimum height = .6cm, minimum width = 2cm,fill=green!30](state1){$5 \times 5$ convolution};
\node[draw,below=.5cm of state1, minimum height = .6cm, minimum width = 2cm,fill=green!30](state2){SELU};
\node[draw,below=.5cm of state2, minimum height = .6cm, minimum width = 2cm,fill=green!30](state3){batch normalization};

\draw[-triangle 60] (state0) -- (state1);
\draw[-triangle 60] (state1) -- (state2);
\draw[-triangle 60] (state2) -- (state3);

\draw [decorate,decoration={brace,amplitude=10pt,raise=4pt},
xshift=20cm] ($(state0)+(1.6,.4)$) -- ($(state3)+(1.6,-.4)$);

\node[below=.2cm of state1](final1){};

\node[draw,right=2.2cm of final1, minimum height = .6cm, minimum width=2cm,fill=blue!30](final){\emph{conv layer}};

\end{tikzpicture}
\caption{Basic building block in our neural network (``\textit{conv layer}''; blue). It consists of (in green) a dropout layer to prevent overfitting, a convolutional layer to convolve the input images, the application of an activation function (i.e., SELU), and a batch normalization layer.} \label{fig:networkblock}
\end{center}
\end{figure}

All neurons have Scaled Exponential Linear Unit (SELU) activation functions %
\citep{Klambauer2017}, except for the last layer which uses the identity function, as usual in regression problems.
This activation function was chosen because it led to better results on the validation set than other possibilities, such as ReLU, leaky ReLU and ELU. 
The size of the convolutional filters is chosen to be 5 $\times$ 5. 
We also add dropout layers to avoid overfitting to the training set, and batch normalization to ensure that the input of each layer is appropriately normalized, facilitating smoother training.
Most layers have stride 1, leading to output images with the same dimensions as the inputs, but we set stride to 2 in some layers in the encoding phase, down-sampling the images at those points.
In the decoding phase, we up-sample the input back to its original size. 
The basic building block for our network considering the above design choices is illustrated in \myfigure~\ref{fig:networkblock}.

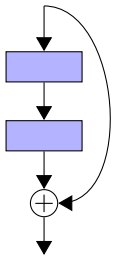
\begin{figure}[t]
\begin{center}
  \begin{tikzpicture}
\node (inputs){};
\node[draw,below=.6cm of inputs, minimum height = .4cm, minimum width = 1cm,fill=blue!30] (state0){};
\node[draw,below=.5cm of state0, minimum height = .4cm, minimum width = 1cm,fill=blue!30](state1){};
\node[draw, circle, inner sep=0pt, minimum size=.2cm, below=.5cm of state1](state2){$+$};
\node[below=.5cm of state2] (state3){};

\draw[-triangle 60] (inputs) -- (state0) node[midway,left]{};
\draw[-triangle 60] (state0) -- (state1) node[midway,left]{};
\draw[-triangle 60] (state1) -- (state2);
\draw[-triangle 60] (inputs.south) to [out=0,in=0] (state2);
\draw[-triangle 60] (state2) -- (state3);
\end{tikzpicture}
\caption{Illustration of a residual connection amongst ``conv'' layers. If the two inputs to the sum have different dimensions, an extra convolutional layer with no activation function is added to the shortcut path before the sum.} 
\label{fig:residualconn}
\end{center}
\end{figure}

Finally, we also use \emph{residual connections} in our network \citep{HeZRS15}.
To construct a residual connection, we take the inputs of a given layer and sum it to the outputs of the layer after that one, as seen in \myfigure~\ref{fig:residualconn}.
In our network, we connect the inputs to the outputs of the second layer, those to the outputs of the fourth layer, and so on.
Residual connections are known to improve the training performance of deep neural networks and were instrumental for recent artificial intelligence breakthroughs, such as AlphaGo Zero \citep{Silver2017}. 
Residual connections have been used in UNets to form ResUNets \citep{KayalibayJS17,Zhang17}. 
We found that the introduction of residual connections dramatically decreased the final output error in the task at hand.

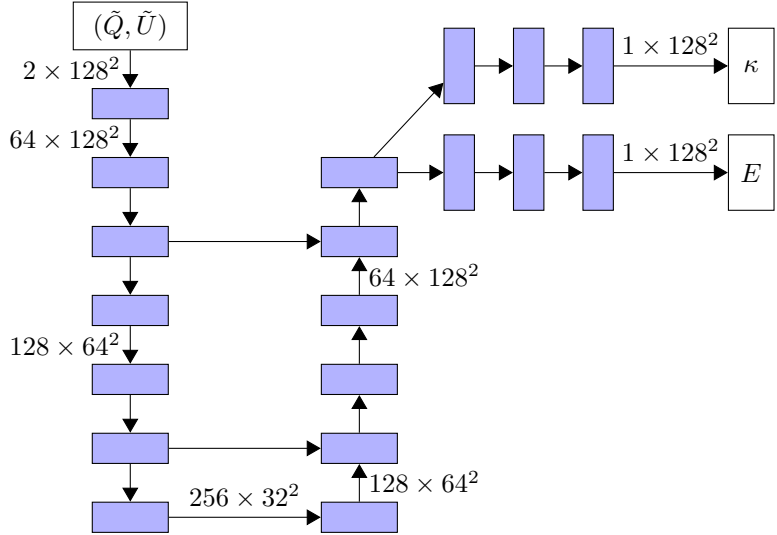
\begin{figure*}[t]
\begin{center}
  \begin{tikzpicture}
\node[draw, minimum height = .6cm, minimum width = 1.5cm] (inputs){$(\tilde{Q},\tilde{U})$};
\node[draw,below=.5cm of inputs, minimum height = .4cm, minimum width = 1cm,fill=blue!30] (state0){};
\node[draw,below=.5cm of state0, minimum height = .4cm, minimum width = 1cm,fill=blue!30](state1){};
\node[draw,below=.5cm of state1, minimum height = .4cm, minimum width = 1cm,fill=blue!30](state2){};
\node[draw,below=.5cm of state2, minimum height = .4cm, minimum width = 1cm,fill=blue!30](state3){};
\node[draw,below=.5cm of state3, minimum height = .4cm, minimum width = 1cm,fill=blue!30](state4){};
\node[draw,below=.5cm of state4, minimum height = .4cm, minimum width = 1cm,fill=blue!30](state5){};
\node[draw,below=.5cm of state5, minimum height = .4cm, minimum width = 1cm,fill=blue!30](state6){};
\node[draw,right=2cm of state6, minimum height = .4cm, minimum width = 1cm,fill=blue!30](state7){};
\node[draw,above=.5cm of state7, minimum height = .4cm, minimum width = 1cm,fill=blue!30](state8){};
\node[draw,above=.5cm of state8, minimum height = .4cm, minimum width = 1cm,fill=blue!30](state9){};
\node[draw,above=.5cm of state9, minimum height = .4cm, minimum width = 1cm,fill=blue!30](state10){};
\node[draw,above=.5cm of state10, minimum height = .4cm, minimum width = 1cm,fill=blue!30](state11){};
\node[draw,above=.5cm of state11, minimum height = .4cm, minimum width = 1cm,fill=blue!30](state12){};
\node[draw,right=.6cm of state12, minimum height = 1cm, minimum width = .4cm,fill=blue!30](stateE0){};
\node[draw,above=.4cm of stateE0, minimum height = 1cm, minimum width = .4cm,fill=blue!30](statek0){};
\node[draw,right=.5cm of stateE0, minimum height = 1cm, minimum width = .4cm,fill=blue!30](stateE1){};
\node[draw,right=.5cm of statek0, minimum height = 1cm, minimum width = .4cm,fill=blue!30](statek1){};
\node[draw,right=.5cm of stateE1, minimum height = 1cm, minimum width = .4cm,fill=blue!30](stateE2){};
\node[draw,right=.5cm of statek1, minimum height = 1cm, minimum width = .4cm,fill=blue!30](statek2){};
\node[draw,right=1.5cm of stateE2, minimum height = 1cm, minimum width = .6cm](stateE3){$E$};
\node[draw,right=1.5cm of statek2, minimum height = 1cm, minimum width = .6cm](statek3){$\kappa$};

\draw[-triangle 60] (inputs) -- (state0) node[midway,left]{$2\times 128^2$};
\draw[-triangle 60] (state0) -- (state1) node[midway,left]{$64\times 128^2$};
\draw[-triangle 60] (state1) -- (state2);
\draw[-triangle 60] (state2) -- (state3);
\draw[-triangle 60] (state3) -- (state4) node[midway,left]{$128\times 64^2$};
\draw[-triangle 60] (state4) -- (state5);
\draw[-triangle 60] (state5) -- (state6);
\draw[-triangle 60] (state6) -- (state7) node[midway,above]{$256\times 32^2$};
\draw[-triangle 60] (state7) -- (state8) node[midway,right]{$128\times 64^2$};
\draw[-triangle 60] (state8) -- (state9);
\draw[-triangle 60] (state9) -- (state10);
\draw[-triangle 60] (state10) -- (state11) node[midway,right]{$64\times 128^2$};
\draw[-triangle 60] (state11) -- (state12);
\draw[-triangle 60] (state12) -- (stateE0);
\draw[-triangle 60] (state12) -- (statek0);
\draw[-triangle 60] (stateE0) -- (stateE1);
\draw[-triangle 60] (statek0) -- (statek1);
\draw[-triangle 60] (stateE1) -- (stateE2);
\draw[-triangle 60] (statek1) -- (statek2);
\draw[-triangle 60] (stateE2) -- (stateE3) node[midway,above]{$1\times 128^2$};
\draw[-triangle 60] (statek2) -- (statek3) node[midway,above]{$1\times 128^2$};
\draw[-triangle 60] (state5) -- (state8);
\draw[-triangle 60] (state2) -- (state11);

\end{tikzpicture}
  \caption{Complete network architecture, with residual connections omitted. Each blue layer contains the components shown in \myfigure~\ref{fig:networkblock}, except for the last conv layers before the outputs to which no activation function is applied. Arrows coming out of a network layer always carry the outputs of that layer, and arrows coming into a layer are always inputs to it. When more than one arrow comes out of a layer, the outputs are duplicated on each arrow. If more than one arrow comes into a layer, the several inputs are concatenated. The shape of the outputs of a layer is omitted if it is equal to the shape of its inputs. The network has just over 5.4 million parameters in total, and the receptive field for each pixel of the output has a size of $101 \times 101$ pixels.} \label{fig:networkarchitecture}
\end{center}
\end{figure*}

The assemblage of the building blocks into our full network is depicted in \myfigure~\ref{fig:networkarchitecture}, where we chose to omit the residual connections.
The ResUNet provides a non-linear mapping between $\mathbb{R}^{2\times 128 \times 128}$ (representing $(\tilde{Q},\tilde{U})$), to $\mathbb{R}^{2\times 128 \times 128}$ (representing $(E,\kappa)$).
The representation in the middle of the bottleneck of the network is an element of $\mathbb{R}^{256 \times 32 \times 32}$, and it will form a processed version of the information in the input maps, optimized for generation of the output maps.

\section{Data}
\label{sec:data}

We use simulated data to develop and compare the gravitational lensing reconstruction algorithms.
We prepare 11200 independent realizations of simulated CMB maps ($\tilde{Q}$, $\tilde{U}$, $E$, $\kappa$), each 5 deg $\times$ 5 deg in size on sky, by extracting 160 patches from each of 70 full-sky maps.
Note that the training, validation, and test sets are selected from separate sets of full-sky maps, so no contamination from large-scale information in the training set is possible.
The images are pixelized into smaller images that are 128 $\times$ 128 pixels using the Lambert azimuthal equal-area projection.
These simulations are created given the $E$ and $\kappa$ power spectra generated based on Planck 2013 best-fit $\Lambda$CDM cosmology: $\Omega_b h^2$ = 0.0222, $\Omega_{\rm CDM} h^2$ = 0.1185, $A_s$ = 2.21 $\times 10^{-9}$, $n_s$ = 0.9624, $\tau$ = 0.0943, $H_0$ = 67.94, using CAMB~\citep{cambpaper} for the theory spectra, and {\sc HEALPix}\footnote{http://healpix.sourceforge.net} for synthesizing the $a_{\ell m}$'s. 
We project the $a_{\ell m}$'s to an equirectangular projection
and lens them with the {\sc quicklens}\footnote{https://github.com/dhanson/quicklens} package. 

\FigureInputmapsnoisev2

We add complexity and realism to the simulations by including various white noise levels of 1, 2, 5 $\mu$K-arcmin to the $(\tilde{Q},\tilde{U})$ maps, a 1 arcmin beam smoothing, and an apodization mask.  Examples of these images with varying noise levels can be seen in \myfigure~\ref{fig:inputmaps}.

\subsection{Data preparation and network optimization}
\label{sec:datapreparation}

The 11200 simulations were separated in $80:10:10$ proportion into training, validation and test sets. 
A different network was trained for each noise level, beam smoothing and mask configuration.
All results presented here come from running a trained network on the test set, which was only used once the architecture had been optimized with respect to its performance on the validation set.
Some deeper architectures than the final one used here were tried with no performance improvement, but we do not claim to have explored the full parameter set.
Training is done using the Adam optimizer on mini-batches of 32 samples, with initial learning rate 0.25 which is halved every time the validation error has not improved for three consecutive epochs. 
The dropout rate is set to 0.3. 
The network is considered to have converged and training is stopped if the validation error does not improve for ten consecutive epochs.
Networks were trained on a single NVIDIA P100 GPU, using Keras with a TensorFlow backend.
Training took roughly 200 seconds per epoch, for a total of three to five hours.
Running the trained network on each sample of the test set takes 11 ms, or a total of about one minute to run over all 1120 realizations in the test set if we include the time to load the network and simulations from memory.

Both inputs and outputs will be images with $128 \times 128$ pixels.
Before training, we calculate the standard deviation of pixel values across all $\tilde{Q}$ maps in the training set,
and normalize all $\tilde{Q}$ inputs to the network in the training, validation, and test sets by this value.
The corresponding normalizing factors are also applied to each of $(\tilde{U}, E, \kappa)$.

We employ mean squared error in image space as the loss function for training.
Since both outputs are normalized to have unit standard deviation, errors on the outputs are equally weighted.
We choose to use the noise spectrum of the output convergence map $\kappa$ defined in \myequation~\eqref{eqn:phinoisedefn} as a metric of the network performance,  
allowing for a direct comparison to the performance of standard methods.
We have tried to introduce loss functions closer to this metric in the training of the network, such as mean squared error in Fourier space, but found no improvement on the performance.

\section{Results}
\label{sec:results}

In this section, we first compare the trained network's output $\hat{E}$ and $\hat{\kappa}$ with the true $E$ and $\kappa$. We then compare the $\kappa$ noise spectra between traditional methods and neural network outputs. We perform null tests of the networks by passing unlensed CMB maps through them.
Finally, we perform checks on the robustness of the neural network approach against different input cosmologies, and use a simple toy scenario to demonstrate how to carry out parameter estimation from the network's results.

\subsection{Recovering unlensed E and \texorpdfstring{$\kappa$}{kappa}}
\label{sec:recover}

We apply the network to the problem of recovering the unlensed CMB maps and the convergence map from observed polarization maps, as described in \S\ref{sec:cmbobservationslensing}. 
We will always take as inputs the lensed $Q$ and $U$ maps, while the desired outputs are the unlensed $E$ map as well as $\kappa$. 

\FigureEmaps

\Figurekmaps

\myfigures~\ref{fig:inputmaps},~\ref{fig:Emaps} and~\ref{fig:kmaps} show an example of the input $(\tilde{Q},\tilde{U})$ maps, the target $(E,\kappa)$ maps, and the predicted $(\hat{E},\hat{\kappa})$ maps from the network for one realization in the test set. 
From visual inspection we can tell that for noiseless inputs, the network recovers structures in both the $E$-mode map and the $\kappa$ map fairly well -- the red and blue clumps trace each other in the true vs. the predicted maps.
The bottom panels show the differences of the predicted maps from the input maps. 
When the inputs are noiseless, both the residuals from the recovered $E$ and $\kappa$ maps
are within a few tens of percent of the input maps' maximum pixel value.
From the residual maps, it is apparent that most large scale structure present in the $\kappa$ map was captured by the network predicted map,
while the residual $E$-mode map has visible structure that is not being captured.
However, noise has a smaller effect on $E$ than on $\kappa$, whose recovery very visibly degrades to the extent that at 5 $\mu$K-arcmin the network can recover structure at only the larger scales. 

To make these observations more precise and to help quantify the efficacy of the network, we compute the power spectrum
of the recovered images and compare them to the true $E$ and $\kappa$, as shown in \myfigure~\ref{fig:spectra}.

\FigureEkspec

\myfigure~\ref{fig:spectra} shows the power spectra of the recovered $E$ and $\kappa$ maps for three input map noise levels: noiseless, 1 $\mu$K-arcmin, and 5 $\mu$K-arcmin.
For the $E$-mode map, the mode recovery gets systematically worse as $\ell$ increases,
but adding noise to the input maps does not degrade $E$-mode recovery significantly from the noiseless case.
For the $\kappa$ map, on the other hand, from noiseless inputs the network is able to recover more than 90\% of the $\kappa$ map fluctuations (80\% in power spectrum) for the entire $L$ range we consider. 
When noise is added to the input maps, the recovery visibly degrades. 
We note that even in the noiseless case, the E-mode recovery is worse than the $\kappa$ recovery both on large angular scales and small angular scales. This is slightly surprising because the mathematical conversion between $(Q,U)$ and $E$ is very simple, compared to that between $(\tilde{Q},\tilde{U})$ and $\kappa$.
This may mean that recovery of maps that have oscillatory amount of correlation across different angular scales is a more challenging problem than that of maps whose correlation across different angular scales is smooth.
One might speculate that since the conversion of $(Q, U)$ to $(E, B)$ is non-local, it will not be straight-forward for the network to recover. To investigate this, we built a network that converts $(\tilde{Q},\tilde{U})$ to $(Q,U)$ and $\kappa$ and found no improvement in the $E$ recovery.
While this is an intriguing problem, in this article we focus on $\kappa$ recovery, so we leave optimizing for $E$ recovery for future work.

One way to compare this performance against that of traditional reconstruction methods is to compare the noise-per-mode in the recovered lensing convergence. To do that, we estimate the equivalent of the noise spectrum in \myequation~\eqref{eqn:phinoisedefn} for the network.
To compare $\hat{\kappa}$ recovered with the network directly to QE-reconstructed $\dbhat{\kappa}$, 
we normalize it by $1/R$ to get the equivalent of the unbiased $\kappa$,
\begin{equation}
\dbhat{\kappa} = \frac{1}{R}\hat{\kappa},
\label{eqn:kappanorm}
\end{equation}
where
\begin{equation}
R = \frac{ \left\langle \kappa \hat{\kappa}^* \right\rangle}  { \left\langle \kappa\kappa^* \right\rangle }
\label{eqn:kappanormratio}
\end{equation}
averaged over the entire validation set.\footnote{There are other ways one can define noise in the $\hat{\kappa}$ maps. For example, it can be extracted through the correlation coefficient of $\kappa$ and $\hat{\kappa}$. We chose to define noise this way to symmetrize $R$ in Eqn.~\eqref{eqn:qenorm} and $R$ in Eqn.~\eqref{eqn:kappanorm}. In ideal conditions (i.e.\ white noise, azimuthally-symmetric filtering), $R$ as defined in~\eqref{eqn:kappanormratio} would give identical results as $R$ as defined in~\eqref{eqn:qenorm}.} 
This is analogous to the response in QE $\phi$ reconstruction~\citep{omori17}, describing how much of the true map is correctly estimated. Note that by construction we will then have
\begin{equation}
\left\langle \kappa \dbhat{\kappa}^* \right\rangle = \left\langle \kappa \kappa^* \right\rangle
\end{equation}
on the validation set.

\FigureclkkQEIMvsResUNet

After normalizing with $R$, we compute the power spectrum of $\dbhat{\kappa}$ and extract the noise spectrum by differencing the auto-spectrum of $\dbhat{\kappa}$ and $\kappa$, as in \myequation~\eqref{eqn:phinoisedefn}.
The noise spectra from $\kappa$ reconstructed through the QE and from the ResUNet are shown in \myfigure~\ref{fig:clkkQEvsResUNet}.
We see that at angular scales below $L$ of 2500, the noise levels from the ResUNet are lower than the QE's noise levels, regardless of the input maps' noise levels.
While 5 $\mu$K-arcmin is a high enough noise level that the QE is in principle close to optimal, the actual reconstruction by the QE on these maps does not provide as low as noise level as the QE analytic $N_0$ estimates, mainly because of E-B mode mixing due to boundary effects.
We choose to highlight 1 and 5 $\mu$K-arcmin as these are realistic next-generation noise levels, but note that a similar performance difference is obtained for noiseless maps.
From these results, we conclude that ResUNets outperform the QE for input map noise levels below 5 $\mu$K-arcmin.

In \myfigure~\ref{fig:clkkIMvsResUNet}, we also show the $N_0$ noise curves from the iterative estimator using the formalism of \citet{smith10} outlined in \S\ref{sec:maximumlikelihood}.
Once again we highlight 1 and 5 $\mu$K-arcmin noise levels, as with noiseless inputs the iterative estimator can reach zero error.
We see that the neural networks provide comparable performance to the iterative
estimator across a wide range of angular scales.
This means that the neural network approach is able to extract information at efficiency close to the iterated EB estimator.
Since the inputs to the network are CMB $(\tilde{Q}, \tilde{U})$ polarization maps, the neural network should also include information from the EE estimator of standard methods. 
In other words, the noise levels in the $\kappa$ maps extracted by the neural network are higher than the combined $N_0$ from iterated EB and EE $N_0$. 
With that said, the iterated EB estimator provides the lowest reconstruction noise among individual QE's for future CMB experiment noise levels, so the neural network $\kappa$ noise being not far off from that demonstrates its viability for beyond-QE $\phi$ reconstruction.
We should also note that once the network has been trained on simulations, it can be applied to real data very quickly when compared to other maximum likelihood methods under development.

\FigureUncertainty

In \myfigure~\ref{fig:uncertainty1}, we seek to quantify the variance of the network output when given simulations from the test set.
To obtain these error bars, we calculate the spectrum from each $\hat{\kappa}$, rescale it by $R$, and subtract the average noise spectrum $N_L^{\kappa\kappa}$ shown in \myfigure~\ref{fig:comparisonstoothermethods}.
We use bins of width 127 for \myfigure~\ref{fig:uncertainty1}.
The error bars represent one-standard-deviation variations from the mean power spectrum of the test set. This calculation provides a measure of the effects of cosmic variance and noise variance between different patches in the test set on the neural network predictions.
For noiseless inputs, the relative uncertainty in the first bin is 33\%. The uncertainty decreases to 13\% at $L \sim 1500$ and 10\% at $L \sim 3000$.
With 1 $\mu$K-arcmin noise, as shown in \myfigure~\ref{fig:uncertainty1}, the relative uncertainty in the first bin remains at 33\%, but the increased $N_L^{\kappa\kappa}$ for higher $L$ brings it up to 33\% at $L\sim 1500$ and 117\% at $L \sim 3000$.
With 5 $\mu$K-arcmin noise, the first bin suffers from the increase in noise, so the relative error bar goes up to 48\%. At $L\sim 1500$, we reach 298\%. Beyond that $L$ the noise shoots up so the reconstruction of $\kappa$ is no longer meaningful.
This variation would form a part of the uncertainty budget when the neural networks are applied to real data.

\subsection{Null test}
A basic test to check that the network encodes a sensible mapping of the input lensed maps to the underlying lensing convergence is to pass unlensed maps through the network and compare the output to the noise spectrum. 
For this network to be useful for cosmology, we need to know that when fed with a map with no lensing, the network recovers a field that is uncorrelated with the convergence. 
Therefore, we feed unlensed versions of the $(Q,U)$ maps (that is, maps with $\kappa=0$) through the network trained on lensed $(\tilde{Q},\tilde{U})$ maps.

As a first test, to check that the network has not overfit the training set, we run unlensed versions of the noiseless $(Q,U)$ maps in the training set through, and calculate the cross-spectrum of the true $\kappa$ (present in the training set, but not applied to the maps we run here) and the output field $\hat{\kappa}$. We obtain
\begin{align}
    \langle \kappa \hat{\kappa}^* \rangle < 10^{-4} \langle \kappa \kappa^* \rangle,
\end{align}
showing no significant amount of overfitting.

Next, in defining the noise proxy as the difference between the auto-spectrum of $\dbhat{\kappa}$ and $\kappa$, we have assumed that we can model the $\dbhat{\kappa}$ as $\dbhat{\kappa}=\kappa + n_\kappa$, where $n_\kappa$ is an uncorrelated noise term.
If this model is accurate, we expect the auto-spectrum of $\dbhat{\kappa}$ maps output when given unlensed $(Q,U)$ maps to be consistent with the noise spectrum $N_L^{\kappa\kappa}$ we used in the previous section. 
In other words, 
\begin{equation}
C_L^{\dbhatssk{\kappa}\dbhatssk{\kappa}}=\langle \dbhat{\kappa}\dbhat{\kappa}^*\rangle = \langle \kappa \kappa^*\rangle + \langle n_\kappa n_\kappa^*\rangle = C_L^{\kappa\kappa}+N_L^{\kappa\kappa}.
\end{equation}
To test this, we run unlensed versions of the input maps $(Q,U)$ in the test set, with noise levels of 1 and 5 $\mu$K-arcmin, through the networks trained on lensed inputs at the same level of noise.
We then rescale the outputs of the network using the same factor $1/R$, and compare $C_L^{\dbhatssk{\kappa}\dbhatssk{\kappa}}$ from unlensed inputs to the noise spectra that we extracted from the tests on lensed inputs. 
For both white noise levels, the $C_L^{\dbhatssk{\kappa}\dbhatssk{\kappa}}$ from unlensed inputs align well with the noise spectrum with percent-level differences, as we can see in \myfigure~\ref{fig:nulltest}.
The difference can be attributed to two potential causes: (1) a subdominant part of the noise $n_\kappa$ that correlates with the lensing convergence field, similar to $N_1$ in standard QE methods; (2) the $R$ computed from the training set fluctuates high/low and biases $\dbhat{\kappa}$.
For future work, it will be important to characterize how the network interacts with higher-order noise terms. 
In conclusion, this check confirms the assumption of $\dbhat{\kappa}=\kappa + n_\kappa$ to be good to within a few percent.

\FigureNullTest

\subsection{Tests on cosmology}
\label{sec:cosmology}

To test that we can apply the network on actual data, we should check whether it will be sensitive to changes in the input $(Q,U)$ maps' cosmology, and has not simply learned to reproduce the cosmology in the training set.
To this end, we give noiseless $(\tilde{Q},\tilde{U})$ maps that are generated with different parameters as inputs to the network trained with the fiducial set of $(\tilde{Q}, \tilde{U}, E, \kappa)$ maps.
The two different cosmologies have $\Omega_{\rm CDM} h^2 = 0.1085$ and $\Omega_{\rm CDM} h^2 = 0.1285$ respectively (while $\Omega_\text{CDM} h^2 = 0.1185$ for the fiducial cosmology), with all the other parameters fixed to the fiducial.
We found that the recovered $\kappa$ spectrum is significantly different from the spectrum recovered from the fiducial set. 

To quantify the difference, we pose the null hypothesis: ``if we apply the network trained on maps generated from the fiducial cosmology on real data with cosmologies different from the fiducial cosmology, we will get the same output as the fiducial cosmology.''
We calculate the $\chi^2$ of each sample of the recovered $\kappa$ spectrum from the two different input cosmologies compared against the average recovered $\kappa$ spectrum from the fiducial cosmology, as
\begin{align} \label{eqn:chisq}
    \chi^2 = (\boldsymbol{d} - \boldsymbol{\mu}) ^{\dagger} \mathbf{C}^{-1} 
(\boldsymbol{d} - \boldsymbol{\mu}),
\end{align}
where $\boldsymbol{d}$ is the binned noise-debiased $\dbhat{\kappa}$ power spectrum, $\boldsymbol{\mu}$ is the $\kappa$ spectrum in the fiducial cosmology, and $\mathbf{C}$ is the covariance matrix that describes the $L$ to $L^{\prime}$ bin covariance. We construct $\boldsymbol{\mu}$ and $\mathbf{C}$ from the outputs of the network for all realizations in the test set (generated with the fiducial cosmology). We note the error bars shown in Fig.~\ref{fig:uncertainty1} are the square roots of the diagonal elements in $\mathbf{C}$. Using $\chi^2$, we rule out the null hypothesis at $2.9\pm 0.9$ and $4.5 \pm 1.4$ $\sigma$ respectively.
This demonstrates that the network is sensitive to differences in the input maps' cosmology. 

To demonstrate that we can use the recovered $\dbhat{\kappa}$ from neural networks for extracting cosmological parameters, we  use the $\dbhat{\kappa}$ spectrum for parameter estimation in a simple case, where we only fit for the parameter $\Omega_{\rm CDM} h^2$. 
We construct a Gaussian likelihood for the parameter $\theta$ = $\Omega_{\rm CDM} h^2$ given the $\dbhat{\kappa}$ spectrum: 
\begin{multline}
\mathcal{L}(\theta| \boldsymbol{d}) \propto  %
\exp \left( -\frac{1}{2} 
(\boldsymbol{d} - \boldsymbol{\mu}(\theta)) ^{\dagger} \mathbf{C}^{-1} 
(\boldsymbol{d} - \boldsymbol{\mu}(\theta)) \right),
\end{multline}
where $\boldsymbol{\mu}(\theta)$ is the $\kappa$ spectrum given different $\Omega_{\rm CDM} h^2$ values while $\mathbf{d}$ and $\mathbf{C}$ are as in \myequation~\eqref{eqn:chisq}, with the covariance constructed from simulations. 
When noise debiasing the $\dbhat{\kappa}$ spectrum, we always subtract the noise spectrum estimated from the fiducial set from the $\dbhat{\kappa}$ spectrum recovered from input $(\tilde{Q}, \tilde{U})$ maps of different cosmologies.

From the set of simulations with the fiducial cosmology, the mean and standard deviation of the distribution of maximum-likelihood $\Omega_{\rm CDM} h^2$ values from 320 realizations are 0.1183$\pm$0.0016, recovering the input $\Omega_{\rm CDM} h^2 = 0.1185$ at within about 0.1$\sigma$.
From the set of simulations that has different input $\Omega_{\rm CDM} h^2$  values, the mean of the distribution of maximum-likelihood $\Omega_{\rm CDM} h^2$ values are 0.1096 for input $\Omega_{\rm CDM} h^2 = 0.1085$ and 0.1260 for $\Omega_{\rm CDM} h^2 = 0.1285$, respectively.
The low $\Omega_{\rm CDM} h^2$ set is biased high, whereas the high $\Omega_{\rm CDM} h^2$ is biased low. 
This is due to the noise debias term being too small for the low $\Omega_{\rm CDM} h^2$ case and too large for the high $\Omega_{\rm CDM} h^2$ case.
In standard cosmology parameter estimation from the lensing power spectrum, since the noise debias terms are cosmology-dependent, one can correct for the difference in the noise terms between the fiducial cosmology and the sampled cosmology to avoid this bias (see e.g. Appendix C of~\citet{planck2015XVlens}).
In our case, when we apply the network to data and use $\dbhat{\kappa}_{\rm data}$ for parameter estimation, we can apply similar corrections by computing the noise terms for many different input cosmologies and obtaining the numerical derivatives of the noise terms with respect to the CMB power spectra. 
But that is outside the scope of current work. 
From this simple test, we conclude that the neural network recovered $\kappa$ map and the input $\Omega_{\rm CDM} h^2$ have a one-to-one mapping.
While more detailed calibration has to be done, there is no conceptual roadblock for parameter estimation from these maps.

\section{Discussion}
\label{sec:discussion}

The results outlined in the previous section provide a strong argument for the viability of using neural networks to reconstruct the $\kappa$ map for next-generation low-noise CMB experiments. In this section, we comment on some of the possible issues and what can be done to mitigate them.

One clear feature of \myfigure~\ref{fig:comparisonstoothermethods} is the $\kappa$ reconstruction noise shooting up for $L \gtrsim 2000$ when the inputs have 5 $\mu$K-arcmin noise.
A similar phenomenon also happens for 2 $\mu$K-arcmin input noise after $L \gtrsim 3000$.
These angular scales are approximately where the RMS of the noise added to the input (signal) maps dominates the signal's RMS, thus submerging information contained in these modes of the inputs.
This issue affects the performance of the neural networks much more sharply than it affects that of traditional methods.
The correlation between the angular scales at which the input's signal-to-noise ratio falls below 1 and the angular scales at which the output $\kappa$ power spectrum degrades leads us to conjecture that the network is using information that is more local in angular scales than standard methods.
 
One way to resolve this would be to impose a physical model on the $\kappa$ power spectrum: for example, one can deduce the smaller angular scale mode information from the structure contained in the larger angular scales (lower-$L$) modes, given our knowledge of the shape of the lensing convergence power spectrum.
However, this is out of the scope of this work where we are interested in a model-free inference of $\kappa$ and primordial E.

One other feature of the results, already apparent in \myfigure~\ref{fig:spectra} even on the noiseless data, is a decrease in recovered power for the low-$L/\ell$ modes in both $\kappa$ and $E$.
This manifests itself in \myfigure~\ref{fig:comparisonstoothermethods} as a sharp increase in the noise levels.
This feature is an effect of the finite size of the maps we have used (5 degrees across).
We have tested this hypothesis by training networks to learn the same task on smaller maps, obtained by cutting out sections of the simulations used in this paper.
We found that the recovered power starts to drop similarly at smaller angular scales (larger $L/\ell$ values) when we decrease the map size. This can be seen in \myfigure~\ref{fig:smallerimages}. 
Extrapolating this tendency, we could improve the results in the low-$L/\ell$ region by performing lensing reconstruction on a larger patch of sky.
We could also increase the receptive field of the networks used, although deeper networks in the same family resulted in no improvement in the validation phase of this work.

\Figuresmallerimages

One final possible issue is the inherent randomness associated with the training of a neural network.
Since the weights are randomly initialized, and the batches used to calculate the gradient at each step are randomly selected from the training set, the final mappings learned by two networks with different initializations will not be exactly the same.
To test how important this effect is, we trained 20 networks with different initializations on the noiseless data, and calculated the power spectra for the $\kappa$ maps predicted by each network for each realization in the test set.

In order to evaluate the spread of predictions from different networks, we calculate the power spectrum $C_{L,i\alpha}^{\kappa\kappa}$ of realization $i$ obtained by network $\alpha$, and bin it in the same way as we did for plotting.
Denoting the binned spectra as $C_{b,i\alpha}^{\kappa\kappa}$, we then evaluate
\begin{equation}
R_{b,i\alpha}=\frac{C_{b,i\alpha}^{\kappa\kappa}}{\left\langle C_{b,i}^{\kappa\kappa} \right\rangle_{\alpha}},
\end{equation}
where $\left\langle C_{b,i}^{\kappa\kappa} \right\rangle_{\alpha}$ is the binned power spectrum for each realization averaged over the results of all 20 networks.
$R_{b,i\alpha}$ are dimensionless quantities whose spread around 1 gives us a measure of the uncertainty due to the randomness of the neural network training algorithm.
We found that the standard deviation of $R_{b,i\alpha}$ over all networks, realizations, and bins was 3.0\%.
If a lower variability is desirable, we could use an ensemble of networks to find a final result, or use weight averaging as introduced in \citet{Izmailov18}.
We leave these refinements to future work.

\section{Conclusion and Outlook}
\label{sec:conclusion}

In this work, we demonstrated that deep learning algorithms (in this work, Residual UNets) can be used to recover the lensing convergence and unlensed CMB maps in simulated data. 
The networks were trained on 5 $\times$ 5 deg$^2$ sized simulated CMB maps.
We first compare our predicted maps (Figs.~\ref{fig:Emaps}, \ref{fig:kmaps}) and power spectra (\myfigure~\ref{fig:spectra}) to the true signals for various noise levels, showing that modes between $\ell=$100 and $\ell=$600 are predicted with errors lower than 10\% for E and 20\% for $\kappa$ for input noise of up to 1 $\mu$K-arcmin.
We then compare our results for the lensing convergence maps to the standard quadratic estimator method at noise levels between 1 and 5 $\mu$K-arcmin.
We show that ResUNets outperform the QE by $50-70\%$ across a wide range of $L$ values in this comparison. This is reflected in the power spectra, and in the ratios of noise spectra in \myfigure~\ref{fig:clkkQEvsResUNet}. In fact, the results approximate maximum likelihood EB estimator results, as we can see in \myfigure~\ref{fig:clkkIMvsResUNet}.

There are some challenges still present in the use of these methods. 
We found that the discrepancies between the true and the recovered maps, even in noiseless cases, increased with the multipole number: small-scale features are more challenging to recover. 
While this is also the case in standard methods, we note that neural networks tend to perform even worse. We speculate that standard methods perform better because of their ability to provide a physical model of the signal and the noise.

In future work, we plan to apply a similar network to recover primordial $B$ modes as well as $E$ and $\kappa$, with more attention paid to parameter estimation from the recovered delensed $E$- and $B$-mode maps.
These are equivalent to delensed CMB polarization maps from standard methods, and will be important for constraining $r$ and $N_{\rm eff}$.
One interesting route would be to directly estimate cosmological parameters from the input CMB maps themselves, as introduced in e.g.~\citet{Ravanbakhsh2017}.
In addition, we plan on including simulations of galactic and extragalactic foregrounds in the input maps to both extract the foreground components and study their effects on $\kappa$ and unlensed CMB recovery.
To extend this network usage for actual data that are often taken from larger regions of the sky, we also need to use simulations from larger sky patches.
This might necessitate the use of different network architectures such as group-equivariant convolutional networks \citep[in particular, the spherical convolutional networks in][]{Cohen2018,Kondor2018}, as the flat-sky approximation will no longer be valid.
During the revision stages of this manuscript, other architectures have been proposed to address this issue~\citep{Perraudin2018, Krachmalnicoff2019}.
It would also be interesting to apply techniques similar to those in this work to the removal of other foregrounds which are hard to model explicitly.
We expect the inherent non-linearities of deep neural networks to be helpful in such tasks. 

The CMB is a potentially powerful data set with which to explore and develop deep learning techniques. 
Because standard techniques to analyze the CMB are quite mature and rich with physical insights, we can develop a better picture of what can be understood with deep learning approaches by comparing the information recovered using neural networks to the standard methods.
This helps us uncover opportunities to improve on standard analyses.
An area where this is especially true is extraction of information contained in the input maps that is not as well-modeled as the CMB itself, such as that coming from galactic foregrounds, instrumental noise, and systematics.

Machine learning can be extremely effective for cosmological data analyses.
However, in order for us to fully leverage its power we first need to elucidate how standard statistical quantities like signal/noise (co-)variance are extracted in each specific application.
This is particularly relevant as machine learning tools are increasingly utilized for scientific analysis and to aid in gaining physical insight.
This work represents a small step in working out one example of connecting standard physical analyses of gravitational lensing to a neural network approach.

\section*{Acknowledgments}

\subsection*{Author Contributions}
\noindent\textbf{Caldeira}: Performed all neural net computational work; innovated the choice of NN architecture; performed all NN diagnostic analysis; kept the work alive and pushed it through the challenging stages.\\
\textbf{Wu}: Directed scientific analysis related to CMB; established the noise measure as a method to compare between QE, max likelihood and NN; performed QE analysis of maps; performed CMB Simulations; performed parameter estimation; guided NN diagnostics; guided analysis.\\
\textbf{Nord}:
Initiated problem concept; initiated NN architecture; performed initial network tests on toy data; guided narrative of paper; guided analysis.\\
\textbf{Avestruz}:
Consultation on analysis methods; writing manuscript.\\
\textbf{Trivedi}:
Consultation on NN methods.\\
\textbf{Story}: Initiated problem concept; performed initial network tests on toy data; generated initial set of CMB simulations.

The authors thank Daniel Gruen for discussions early on. We thank Wayne Hu for useful discussions. We thank earlier paper reviewers, Ren\'ee Hlo\v{z}ek, Alessandro Manzotti, and Eyj\'olfur Guðmundsson for their comments. 

This work is supported by the Deep Skies Community (\url{deepskieslab.com}), which helped to bring together the authors and reviewers. The authors of this paper have committed themselves to performing this work in an equitable, inclusive, and just environment, and we hold ourselves accountable, believing that the best science is contingent on a good research environment.

This manuscript has been authored by Fermi Research Alliance, LLC under Contract No.\ DE-AC02-07CH11359 with the U.S. Department of Energy, Office of Science, Office of High Energy Physics.

JC was supported in part by NSF Grant No.\ PHY-1720480.  WLKW, BN, CA were supported in part by the Kavli Institute for Cosmological Physics at the University of Chicago through an endowment from the Kavli Foundation and its founder Fred Kavli. 
CA was in addition supported by the Kavli Institute for Cosmological Physics at the University of Chicago through NSF Grant No.\ PHY-1125897 and the Enrico Fermi Institute. ST was supported by the National Science Foundation under Grant No.\ DMS-1439786 while the author was in residence at the Institute for Computational and Experimental Research in Mathematics in Providence, RI, during the non-linear algebra program. 

\newpage
\bibliographystyle{model2-names}%
\bibliography{iclr2017_conference}

\end{document}